\documentclass[useAMS,usenatbib]{mn2e}
\usepackage{graphicx}
\usepackage{latexsym}
\usepackage{psfig}
\usepackage{subfigure}
\usepackage{url}

\title{The fragmentation of expanding shells I:\\Limitations of the thin--shell approximation}

\author[James E. Dale, Richard W\"unsch, Anthony Whitworth, Jan Palou\v{s}]{James E. Dale$^{1}$\thanks{E-mail: jim@ig.cas.cz (JED)},
Richard W\"unsch$^{1,2}$,Anthony Whitworth$^{2}$, Jan Palou\v{s}$^{1}$\\
$^{1}$ Astronomical Institute, Academy of Sciences of the Czech Republic, Bocni II 1401/2a, 141 31 Praha 4\\
$^{2}$School of Physics and Astronomy, Cardiff University, Queens Buildings, The Parade, Cardiff, CF24 3AA}

\begin{document}

\pagerange{\pageref{firstpage}--\pageref{lastpage}} \pubyear{2006}

\maketitle

\label{firstpage}

\def\mnras{MNRAS}
\def\apj{ApJ}
\def\aap{A\&A}
\def\apjl{ApJL}
\def\apjs{ApJS}
\def\bain{BAIN}
\def\araa{ARA\&A}
\def\pasp{PASP}
\def\aj{AJ}
\def\ga{\sim}

\begin{abstract}
We investigate the gravitational fragmentation of expanding shells in the context of the linear thin--shell analysis. We make use of two very different numerical schemes; the FLASH Adaptive Mesh Refinement code and a version of the Benz Smoothed Particle Hydrodynamics code. We find that the agreement between the two codes is excellent. We use our numerical results to test the thin--shell approximation and we find that the external pressure applied to the shell has a strong effect on the fragmentation process. In cases where shells are not pressure--confined, the shells thicken as they expand and hydrodynamic flows perpendicular to the plane of the shell suppress fragmentation at short wavelengths. If the shells are pressure--confined internally and externally, so that their thickness remains approximately constant during their expansion, the agreement with the analytical solution is better.
\end{abstract}


\section{Introduction}
Expanding shells or bubbles are an ubiquitous feature of the interstellar medium (ISM) of the Milky Way and of other galaxies. Radio surveys of the Northern \citep[]{1979ApJ...229..533H,1984ApJS...55..585H} and Southern skies \citep{2002ApJ...578..176M} have detected many tens of shell--like structures, in HI emission, and \cite{2005A&A...437..101E} have used an automated search algorithm to find more than 600 in the outer Milky Way within the Leiden--Dwingeloo HI survey. They calculate that the three--dimensional filling factor (defined as the volume fraction of the Galaxy occupied) of such shells must be at least $5\%$. HI shells tend to be large (10--1000 pc) and correspondingly old (1--10 Myr) objects. Recently \cite{2006ApJ...649..759C} and \cite{2007ApJ...670..428C} report the detection of $\sim600$ infrared shells using the GLIMPSE survey. Infrared shells are revealed by dust and PAH emission and are smaller (0.1--10 pc) and younger ($<1$Myr).\\
\indent \cite{2006ApJ...649..759C} find that the scale height of their sample of bubbles is $\sim60$pc, less than that of supernova remnants and much less than that of planetary nebulae, but very similar to the scale height of HII regions. They show that the bubbles are likely connected to very young OB stars. They find that $\sim25\%$ of their bubbles contain known HII regions, and $\sim13\%$ contain open clusters, whereas they find only a few to be connected with known supernova remnants and none with known planetary nebulae. They suggest that the $\sim75\%$ of bubbles not associated with HII regions may be powered by stars later than B4 which do not have strong enough ionizing fluxes to produce detectable HII regions, and conclude that the majority of infra--red bubbles are driven by feedback from OB stars.\\
\indent Such shells have a profound effect on the dynamics of the ISM and, in particular, may be responsible for triggering star formation. Numerous authors \citep[e.g.][]{2003A&A...408L..25D,2006A&A...446..171Z,2006A&A...458..191D,2008A&A...482..585D} find evidence for populations of young (and often massive)  stars in the peripheries of bubbles, and \cite{2007ApJ...670..428C} find that $\sim18\%$ of their sample of 269 bubbles in the inner Milky Way show evidence of triggered star formation. However, \cite{2007ApJ...670..428C} mention only the triggering mechanism of the bubble overrunning pre-existing clumps in the ISM and causing them to collapse. They do not consider the possibility of the shells themselves becoming gravitationally unstable and collapsing, even though they observe that bubbles often have filamentary structure.\\
\indent Numerous authors have studied the gravitational fragmentation of expanding shells using the thin--shell approximation, \citep[e.g.][]{1983ApJ...274..152V,1994ApJ...427..384E,1994A&A...290..421W,1994MNRAS.268..291W,2001A&A...374..746W,2004MNRAS.351..585P}, and we will not give a detailed disquisition here. The analysis consists of studying the stability of perturbations of a given angular size in the surface density of an infinitesimally--thin shell and hence deriving a dispersion relation, i.e. an expression for the rate of growth $\omega$ of a perturbation as a function of its angular size $\eta$. It is customary to denote the angular size of perturbations by $\eta$, but we break with this convention and use $l$ instead since our analysis consists of decomposing the surface density perturbations into spherical harmonics, where each spherical harmonic mode is characterised by the coefficients $l$ and $m$ ($\eta$ and $l$ may be used interchangeably for describing structures of a given angular size).\\
\indent The thin--shell dispersion relation has the form
\begin{eqnarray} 
\omega(l)=-\frac{3V}{2R}+\sqrt{\frac{V^{2}}{4R^{2}}-\frac{c_{s}^{2}l^{2}}{R^{2}}+\frac{2\pi G\Sigma l}{R}},
\label{eqn:disp_rel}
\end{eqnarray}
where $V$, $R$ and $\Sigma$ are respectively the instantaneous velocity, radius and surface mass density of the shell, $c_{s}$ is the sound speed within the shell. This dispersion relation applies in the case of a shell expanding in a vacuum and is slightly different from that appropriate for a shell expanding in a constant--density medium \citep{1994ApJ...427..384E}.\\
\indent Since, $\omega$ is the perturbation growth rate, a positive value of $\omega$ implies a growing perturbation, a purely real negative value implies dissolution and a complex value indicates the damped oscillation of a perturbation.\\
\indent Despite the heavy reliance of theoretical astrophysics on numerical simulations, papers presenting comparisons of different numerical schemes, or even of different codes using the same scheme are still relatively rare \citep[e.g.][]{1986ApJ...305..281D,1993A&A...272..430D,1997MNRAS.288.1060B,2005ApJS..160....1O,2006MNRAS.370..529D,2008A&A...482..371C, 2008arXiv0810.4599K}. In this paper, we compare not only two different codes, but two radically different numerical schemes, namely the Eulerian Adaptive Mesh Refinement (AMR) method (represented by the FLASH code, \cite{2000ApJS..131..273F}), and the Lagrangian Smoothed Particle Hydrodynamics (SPH) technique (represented by a variant of the code described in \cite{1990nmns.work..269B}). The gravitational fragmentation of a thin shell is an ideal test of the two codes since its relative simplicity allows the derivation of a detailed analytic solution. Despite this simplicity, it is still a problem of considerable astrophysical significance, particularly in the field of triggered star formation. This allows us to simultaneously test the two numerical methods and to test the results against the predictions of the thin--shell approximation and thus to explore its relevance to realistic problems.\\
\indent The outline of the paper is as follows: in section 2, we describe our numerical methods. In section 3, we discuss in detail and justify our choice of initial conditions. Section 4 contains the results of our comparative study, while in Section 5 we discuss the astrophysical implications of this work. We draw our conclusions in Section 6.\\
\section{Numerical Methods}
The purpose of this paper is twofold. Our primary aim is to test the reliability of the thin--shell approximation in realistic astrophysical situations. Our methods involve sophisticated numerical hydrodynamics codes and we make use of two very different computational schemes, largely to ensure that our results are robust, but also with the secondary aim of testing the two codes against each other.\\
\indent There are essentially two ways of representing a fluid numerically. Eulerian schemes impose a stationary mesh on the simulation volume which can be refined at will to ensure adequate resolution of high--density features or regions where there are strong gradients in physical quantities, or made coarser in regions where such gradients are shallow -- such a technique is termed Adaptive Mesh Refinement. Conversely, the Lagrangian Smoothed Particle Hydrodynamic method represents the fluid by discrete objects which can move anywhere under the influences of the forces acting upon them, and whose dimensions are automatically adjusted to give the highest resolution to the densest regions of the gas.\\
\indent We choose to use the publicly available FLASH code version 2.5 \citep{2000ApJS..131..273F}. It uses the block-structured AMR technique implemented in the PARAMESH library \citep{macneice}. Its hydrodynamic module uses the piecewise parabolic method \citep{1984JCoPh..54..174C} which incorporates a Riemann solver to compute fluxes between individual cells. The code is parallelized using the MPI library\footnote{The MPI library can be downloaded from \url{http://www.mcs.anl.gov/research/projects/mpi/} where papers discussing MPI and its applications may also be found.}. Self gravity is included using a fast tree algorithm to be described in a forthcoming paper.\\ 
\indent Our SPH code is a variant of that described in \cite{1990nmns.work..269B} but more recently updated and described in \cite{1995MNRAS.277..362B}. The fluid equations are solved using the SPH formalism, including the standard artificial viscosity prescription described in \cite{1992ARA&A..30..543M}, with $\alpha=1$, $\beta=2$. Self--gravity is implemented by means of a binary tree and very high density regions can be replaced by point--mass sink particles, as described in \cite{1995MNRAS.277..362B}, although we do not make use of this facility in this work, since formation of sink particles indicates non--linear behaviour and we are concerned with testing a theoretical model close to the linear regime. We will go on to study non--linear fragmentation and to study in detail the mass spectrum produced in later papers.\\
\indent To ensure a faithful comparison between the two codes, we analyse their output using a common technique. We are studying the formation of structures in expanding shells, which is essentially a two--dimensional problem. We therefore project our shells onto the surface of a sphere at a sequence of timesteps in our simulations. These two--dimensional surface density maps are then decomposed into spherical harmonics using the publicly--available \texttt{nfft} Fast Fourier Transform library\footnote{The \texttt{nfft} library can be downloaded from \url{http://www-user.tu-chemnitz.de/~potts/nfft/} }.\\
\indent A function $f$ on the surface of a sphere may be represented by
\begin{eqnarray}
f(\theta,\phi)=\sum_{l=0}^{\infty}\sum_{m=-l}^{m=+l}a_{l}^{m}Y_{l}^{m}(\theta,\phi),
\end{eqnarray}
where the $Y_{l}^{m}(\theta,\phi)$ are spherical harmonic functions, the $a_{l}^{m}$ are the associated coefficients and the angular wavenumber $l$ is given by
\begin{eqnarray}
l=\frac{2\pi R(t)}{\lambda},
\end{eqnarray}
where $\lambda$ is the wavelength of a given structure. It is customary to work in terms of the angular power spectrum $C_{l}$, defined by 
\begin{eqnarray}
C_{l}=\frac{1}{2l+1}\sum_{m=-l}^{m=+l}|a_{l}^{m}|^{2}
\end{eqnarray}
but we are interested in the surface density perturbations themselves, which are described by $\sqrt{C_{l}}$. We numerically differentiate the timeseries of $\sqrt{C_{l}}$ produced by the \texttt{nfft} library to determine the growth rate of structure as a function of $l$. We can then directly compare our numerically--generated dispersion relation $\omega_{\rm num}(l)={\rm d ln}(\sqrt{C_{l}})/{\rm d}t$ with the analytic $\omega_{\rm anl}(l)$ given in Equation \ref{eqn:disp_rel}.
\section{Simulations}
\subsection{Avoiding instabilities}
The primary aim of this paper is to study the \textit{gravitational} fragmentation of expanding spherical shells. These shells are usually envisaged as being driven internally by an expanding HII region, wind bubble or supernova bubble, and expanding into an external medium. The internal driving and the external medium introduce two major instabilities, respectively the Rayleigh--Taylor and Vishniac instabilities, which may complicate the evolution of the shell and which we have been careful to avoid. We have done this partly to ensure that our comparison with the thin--shell approximation is pure and involves only the gravitational instability, and partly because of the extreme difficulty of simulating such instabilities on a global scale.\\
\indent Internal driving of the shell may introduce the Rayleigh--Taylor instability, which results from a low--density fluid accelerating a high--density fluid. This leads to mixing at the interface and the descent of `fingers' of dense fluid into the rarefied fluid. Transverse flows between the two fluids then rapidly develop and the Kelvin--Helmholtz instability then also appears at the interface. The flow then becomes extremely complex and difficult to simulate. To avoid the development of the Rayleigh--Taylor instability, the shell must be allowed to evolve in freefall with equal pressures on the inner and outer surfaces of the shell.\\
\indent The sweeping up of an external medium can lead to a second instability described by \cite{1983ApJ...274..152V}. The Vishniac instability results from the interplay between the ram--pressure of the material being collected by the shell, which acts on the outer face of the shell and is always directed radially inwards towards the shell centre, and thermal pressure acting on the inner face of the shell, which is always locally perpendicular to the shell surface. The different geometries of these pressures drive transverse flows in the shell and may accelerate fragmentation. To eliminate the Vishniac instability, the ram pressure of the external medium must be negligible in comparison to the the thermal pressure within the shell itself.\\
\indent This study focuses on shells of fixed mass expanding in a vacuum, or into a medium of such low density that the ram pressure exerted and the mass collected are negligible.\\
\subsection{Radial density profile of the shell}
\begin{figure}
\includegraphics[width=0.45\textwidth]{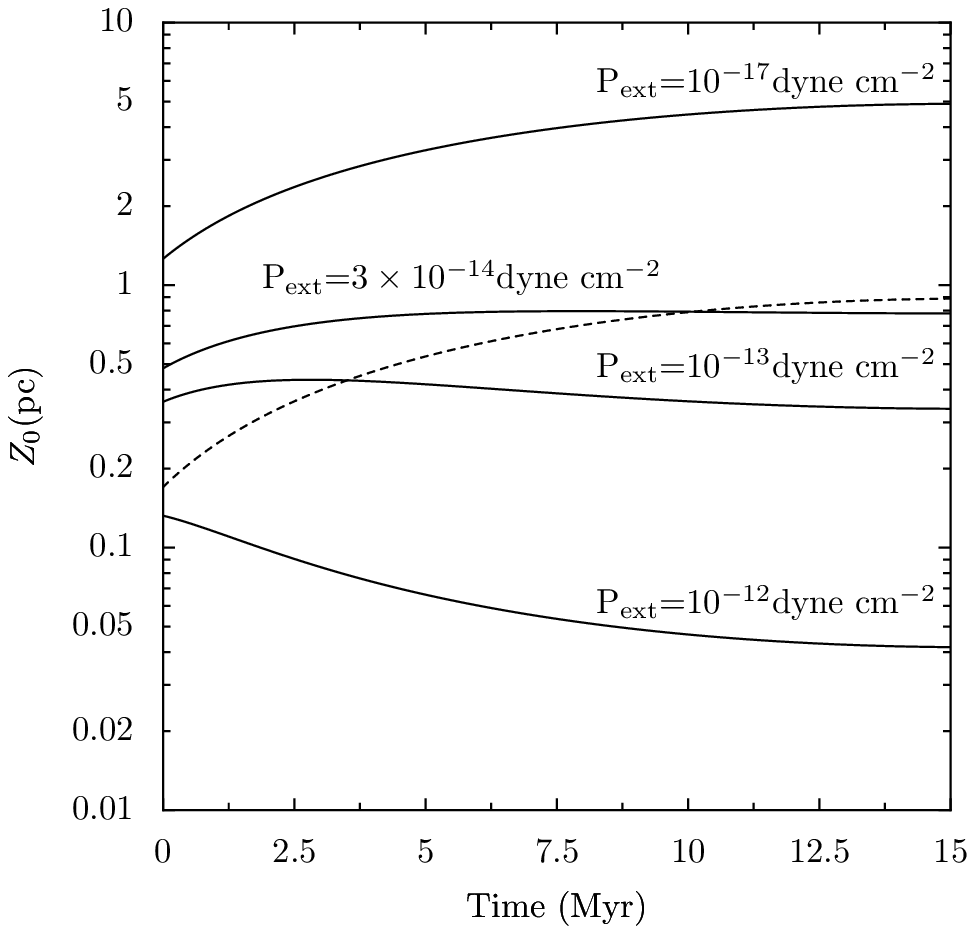}
\caption{Evolution with time of the shell thickness $Z_{0}$ for four values of the external confining pressure (solid lines). For comparison, we also plot the evolution of the shell half--thickness (taken to be  the thickness at which the density $\rho(z)$ falls to half its maximum value) for the $P_{\rm ext}=1\times10^{-17}$ dyne cm$^{-2}$ case (dashed line).}
\label{fig:z0}
\end{figure}

We adopt a shell radial density profile such that the shell is in instantaneous hydrostatic equilibrium at any given radius. Following \cite{1942ApJ....95..329S}, the profile in the $z$--direction of a self--gravitating layer extending to infinity in the $x$-- and $y$-- directions and centred at $z=0$ may be derived by the equation of hydrostatic equilibrium
\begin{eqnarray}
\rho\frac{d\Phi}{dz}+c_{\rm s}^{2}\frac{d\rho}{dz}=0
\end{eqnarray}
with the Poisson equation
\begin{eqnarray}
\frac{d^{2}\Phi}{dz^{2}}=4\pi G \rho.
\end{eqnarray}
These equations yield 
\begin{eqnarray}
\rho(z)=\rho_{0}{\rm sech}^{2}\left[\left(\frac{2\pi G\rho_{0}}{c_{\rm s}^{2}}\right)^{\frac{1}{2}}z\right],
\label{eqn:rho_z}
\end{eqnarray}
where $\rho_{0}$ is the density at $z=0$. To determine this density (and thus the density everywhere), we can integrate through the slab to find the surface density:
\begin{eqnarray}
\Sigma_{z_{0}}=\left(\frac{2c_{\rm s}^{2}\rho_{0}}{\pi G}\right)^{\frac{1}{2}}{\rm tanh}\left[\left(\frac{2\pi G \rho_{0}}{c_{\rm s}^{2}}\right)^{\frac{1}{2}}z_{0}\right].
\label{eqn:sigma_z0}
\end{eqnarray}
Letting $z_{0}\to\infty$ yields
\begin{eqnarray}
\rho_{0}=\frac{\pi G\Sigma_{0}^{2}}{2c_{\rm s}^{2}}.
\end{eqnarray}
The above analysis is valid in plane--parallel geometry. In spherical geometry with $r$ being the radial coordinate and $R$ being the radius of the densest part of the shell, if the thickness of the shell is small compared with its radius, we may set $z=r-R$ everywhere in the shell and use Equation \ref{eqn:rho_z} to calculate the radial density profile of the shell if the shell has a finite thickness. The shells described by Equation \ref{eqn:rho_z} are formally infinitely thick, since the density only falls to zero at infinity. However,  we consider cold shells which are bounded inside and outside by hot rarefied gas exerting a finite (although in some simulations, negligibly small) pressure on the shell. We therefore truncate the density profile of the cold gas, giving the shell a finite initial half--thickness $Z_{0}$, such that the cold gas and hot gas are in pressure equilibrium at $z=\pm Z_{0}$. The relation between the external pressure $P_{\rm ext}$ and the value of $Z_{0}$ follow from Equation \ref{eqn:rho_z}:
\begin{eqnarray}
P_{\rm ext}=\rho(Z_{0})c_{\rm s}^{2}=\rho_{0}c_{\rm s}^{2}{\rm sech}^{2}\left[\left(\frac{2\pi G\rho_{0}}{c_{\rm s}^{2}}\right)^{\frac{1}{2}}Z_{0}\right],
\label{eqn:pext}
\end{eqnarray}
recalling that $c_{s}$ denotes the sound speed in the cold gas that makes up the shell. Rearranging Equation \ref{eqn:sigma_z0} gives
\begin{eqnarray}
Z_{0}=\left(\frac{c_{\rm s}^{2}}{2\pi G \rho_{0}}\right)^{\frac{1}{2}}{\rm tanh}^{-1}\left[\Sigma_{Z_{0}}\left(\frac{\pi G}{2c_{\rm s}^{2}\rho_{0}}\right)^{\frac{1}{2}}\right].
\label{eqn:z0}
\end{eqnarray}
Inserting Equation \ref{eqn:z0} into Equation \ref{eqn:pext} and using ${\rm sech}^{2}\left[{\rm tanh}^{-1}(x)\right]=1-x^{2}$ yields
\begin{eqnarray}
P_{\rm ext}=\rho_{0}c_{\rm s}^{2}\left(1-\frac{\pi G \Sigma_{Z_{0}}^{2}}{2c_{\rm s}^{2}\rho_{0}}\right)
\end{eqnarray}
and finally
\begin{eqnarray}
\rho_{0}=\frac{P_{\rm ext}}{c_{\rm s}^{2}}+\frac{\pi G\Sigma_{Z_{0}}^{2}}{2c_{\rm s}^{2}}.
\end{eqnarray}
This expression for $\rho_{0}$ may be inserted into Equation \ref{eqn:z0} to produce an expression for the shell thickness in terms of the external pressure and the surface density:
\begin{eqnarray}
Z_{0}&=&\left(\frac{c_{\rm s}^{4}}{\pi G\left[2P_{\rm ext}+\pi G\Sigma_{Z_{0}}^{2}\right]}\right)^{\frac{1}{2}}\times \nonumber \\
&&{\rm tanh}^{-1}\left[\left(\frac{\pi G\Sigma_{Z_{0}}^{2}}{2P_{\rm ext}+\pi G\Sigma_{Z_{0}}^{2}}\right)^{\frac{1}{2}}\right].
\end{eqnarray}
\indent Since the shell expands and then contracts, $\Sigma_{Z_{0}}(t)=M/(4\pi R(t)^{2})$. In Figure \ref{fig:z0}, we plot the variation with time of $Z_{0}$ for a shell with the parameters given above and four different values of the external pressure. In the high--pressure case ($P_{\rm ext}=10^{-12}$dyne cm$^{-2}$), the shell initially becomes thinner as it expands, the thickness reaching a minimum value of $\sim0.05$pc. This shell would be very difficult to resolve, requiring a prohibitively large number of SPH particles or many levels of mesh refinement. Conversely, in the case of a confining pressure $P_{\rm ext}=3\times 10^{-14}$dyne cm$^{-2}$, the shell thickens considerably during its evolution, particularly in the first $5$Myr. We therefore choose an intermediate value of the the external pressure, $P_{\rm ext}=10^{-13}$dyne cm$^{-2}$ for which the shell is sufficiently thick to be resolvable and where the thickness of the shell is approximately constant for the duration of its evolution.\\
\subsection{Our initial conditions}
\indent In order to avoid the Rayleigh--Taylor and Vishniac instabilities, we study the momentum--conserving expansion of a shell of fixed mass $2\times10^{4}$M$_{\odot}$ whose initial radius is $10$pc and whose initial outward radial velocity is $2.2$km s$^{-1}$, chosen so that the shell will expand to radius of $\approx23$ pc in a time of $\approx15$Myr, before beginning to contract again. This velocity is rather low in comparison to those observed in real shells, but was selected to ensure that the kinetic energy associated with the expansion was less than the shell's gravitational binding energy, and to keep the ratio of the shell's maximum radius to its initial radius small so that adequate numerical resolution could be maintained during the whole duration of the shell's evolution. We treat the gas in the shell isothermally and set its temperature to 10K. The shell is purely momentum--driven, being powered only by its initial outward radial velocity. We have calculated six models which differ in the external pressure applied to the shell and in the perturbations of density and transverse velocity imposed. In all cases, the pressures within the void inside the shell and in the space outside the shell are equal, so that there is no acceleration of the shell due to these pressures.\\
\indent We consider two cases which we refer to as non--pressure--confined and pressure--confined. In an AMR simulation, it is not possible to simulate regions of true vacuum, so in all of our FLASH simulations, the voids inside and outside the shell are in fact filled with very low density gas. In the non--pressure--confined AMR simulations, the \textit{thermal} pressure, $P_{\rm ext}=10^{-17}$dyne cm$^{-2}$, exerted on the shell by this gas is very low and never becomes dynamically important -- we neglect it entirely in the corresponding SPH simulations. In the pressure--confined AMR simulations, the temperature of the low--density gas inside and outside the shell is higher, so that the $P_{\rm ext}=10^{-13}$ dyne cm$^{-2}$, which ensures that the thickness of the shell stays roughly constant as it expands. In the corresponding SPH calculations, an `external' pressure term is subtracted from the pressure of each SPH particle when calculating pressure gradients (so that particles whose internal pressure is less than the external pressure effectively exert a negative pressure on their neighbours). In all cases (non--pressure--confined and pressure--confined), the pressures inside and outside the shell are kept equal and constant, so that the shell feels no acceleration, and the density of the hot gas used in the AMR calculations is so low that the \textit{ram} pressure it exerts is neglible in comparison with its thermal pressure.\\
\indent In order to study the growth of particular modes, we introduce monochromatic spherical harmonic perturbations $\Sigma_{\rm pert}(\theta,\phi)$ with an amplitude of $10\%$ of the mean shell surface density, and a correlated velocity field derived from \cite{2001A&A...374..746W}. We chose to study the $(l,m)=(12,6)$ and $(35,17)$ modes. The relative amplitude of the surface density perturbation was chosen to be 0.1 to ensure that significant fragmentation occurs before the shell reaches its maximum radius and begins to contract. The thin--shell analysis predicts that these modes should grow. We also study shells seeded with white noise and compare the dispersion relations generated to that predicted by the thin--shell analysis in this case. The parameters of our simulations are given in Table \ref{tab:init}.\\
\indent In SPH simulations of the shells with monochromatic perturbations, it was necessary to reduce the noise intrinsic to the particle distribution, since the level of the noise in our initial conditions was comparable to the amplitude of the perturbations we wished to insert. If the SPH particles are randomly distributed in the shell, the intrinsic noise level in the surface density is $\approx\sqrt{N}/N$, where $N$ is the average number of particles through which a line of sight extending to infinity from the centre of the shell must pass. This may be estimated as $N\approx N_{\rm part}\pi(2\langle h \rangle)^{2} /(4\pi R^{2})$, where $N_{\rm part}$ is the total number of particles in the simulation and $\langle h \rangle$ is the mean smoothing length (note that SPH particles have a radius of $2h$). The value of $\langle h \rangle$ depends on the number of particles and how the particles are distributed radially, i.e. on the thickness of the shell and on its density profile. Using $1.2\times 10^{6}$ particles to simulate shells of the radii and density profiles described in the preceding section, we find that $\langle h \rangle\approx 0.1$pc, so that $\sqrt(N)/N\approx 0.09$. Since this is very similar to the amplitude of the monochromatic perturbations we wished to introduce in the shell, we took steps to reduce the noise level.\\
\indent Distributing the particles on a uniform grid (e.g. a hexagonal close--packed array) can lead to the appearance of grid artefacts. This is not usually the case when the shell is sweeping up additional SPH particles, since the swept--up particles tend to disrupt the uniform grid in the shell and wash out such artefacts, but it is a problem in the simulation of our constant--mass shells. We therefore used an iterative procedure to reduce the noise level in a shell composed of randomly--distributed particles by adjusting the particle masses. For each particle $i$, we drew an imaginary line from the shell centre to infinity through the centre of the particle and generated a list of all the other particles through which this line passed. By integrating through the smoothing kernels of these particles (including particle $i$), we obtain the real surface density of the shell at the angular position of the particle $i$, $\Sigma_{\rm real}^{i}(\theta_{i},\phi_{i})$. We then compare this to the desired perturbed surface density at the position of particle $i$, $\Sigma_{\rm pert}^{i}(\theta_{i},\phi_{i})$ and multiply the mass of particle $i$ by the ratio $\Sigma_{\rm pert}^{i}/\Sigma_{\rm real}^{i}$. In order to avoid numerical difficulties arising from having a large range of particle masses, we impose the further constraint that no particle's mass may differ from the mean particle mass by more than a factor of three, so that the total range of particle masses is less than ten. Since each particle contributes to $\approx N$ lines of sight, this procedure must be repeated iteratively for all particles until the discrepancy between the real and desired surface densities is everywhere within some tolerance (we chose $0.94<\Sigma_{\rm pert}^{i}/\Sigma_{\rm real}^{i}<1.06$). Note that it was not necessary to take such steps in the AMR simulations of monochromatically--perturbed shells, since the amplitude of the intrinsic noise in the AMR density field is very much lower than the $10\%$ perturbation amplitude.\\
\indent In order to study fragmentation at all angular scales simultaneously, we also performed simulations in which white noise was introduced in the initial conditions. We constructed these by retaining the intrinsic noise in the SPH particle distribution and introducing a three--dimensional Maxwellian random velocity field everywhere in the shell, with the peak of the velocity distribution occurring at the sound speed, $c_{\rm s}$. Since $c_{\rm s}<<V$ for most of the shell's evolution, this causes no gross distortions of the shell surface. The presence of this noise in our simulations can be justified on physical grounds by noting that no real expanding shell will be uniform and, even if it were, the instabilities which we have so carefully avoided in these calculations would rapidly introduce inhomogeneities.\\
\indent To ensure a legitimate comparison between the simulations of the shells with noisy initial conditions with the two codes, we interpolated the SPH density and velocity fields onto a grid which was then used as the initial conditions for the corresponding FLASH simulations. We used an SPH particle number of $1.2\times10^{6}$ and grid resolution of $640^{3}$ at the highest refinement level (which contains the whole shell) to give the codes the same linear resultion of $\approx0.1$ pc\\
\indent Our three--dimensional simulations are performed in a Cartesian coordinate system in which the centre of the shell is at $(x,y,z)=(0,0,0)$ and the $z$--coordinate reverts to its usual meaning and should not be confused with the $z$--coordinate defined in Section 3.2.\\
\begin{table*}
\begin{tabular}{lccc}
Perturbation type & $\Sigma_1/\Sigma_0$ & $v_1$ & $P_\mathrm{ext}$ \\
\hline
$l=12$, $m=6$  & 0.1  &  $66$~m\,s$^{-1}$ & $10^{-17} (0)$~dyne\,cm$^{-2}$ \\
$l=35$, $m=17$ & 0.1  &  $32$~m\,s$^{-1}$ & $10^{-17} (0)$~dyne\,cm$^{-2}$ \\
white noise / Maxwell velocity    & 0.1 & $200$~m\,s$^{-1}$ & $10^{-17}
(0)$~dyne\,cm$^{-2}$ \\
\hline
$l=12$, $m=6$  & 0.1  &  $66$~m\,s$^{-1}$ & $10^{-13}$~dyne\,cm$^{-2}$ \\
$l=35$, $m=17$ & 0.1  &  $32$~m\,s$^{-1}$ & $10^{-13}$~dyne\,cm$^{-2}$ \\
white noise / Maxwell velocity    & 0.1 & $200$~m\,s$^{-1}$ &
$10^{-13}$~dyne\,cm$^{-2}$ \\
\end{tabular}
\caption{Initial conditions for our simulations. The first column details the form of the perturbations in the shell, either a monochromatic spherical harmonic perturbation or white noise. The secod column gives the relative amplitude of the surface density perturbations and the third gives the amplitude (for the monochromatic runs) or dispersion (for the white noise runs) of the velocity perturbations. The fourth column gives the pressure applied to the shell (zero for the first three SPH runs).}
\label{tab:init}
\end{table*}
\section{Results}
We study the evolution of shells which sweep up no mass, but with two very different boundary conditions. We first describe simulations in which there is negligible pressure inside or outside the shell to confine it. As the shell expands (i.e. as its radius increases), it also becomes thicker. This is partly due to the decrease in surface density and self--gravity due to the geometrical dilution of the shell, and partly due to the pressure gradients driving material away from the dense core of the shell. We find that this thickening has an important influence on the fragmentation of the shell in that it suppresses fragmentation at shorter wavelengths. This represents a restriction on the validity of the thin--shell approximation. To determine if agreement with the thin--shell analysis can be achieved by preventing the shell thickening, we perform a second set of simulations in which we apply a much stronger pressure to the inner and outer faces of the shell, such that its thickness remains approximately constant during its evolution.\\
\begin{figure*}
\includegraphics[width=0.98\textwidth]{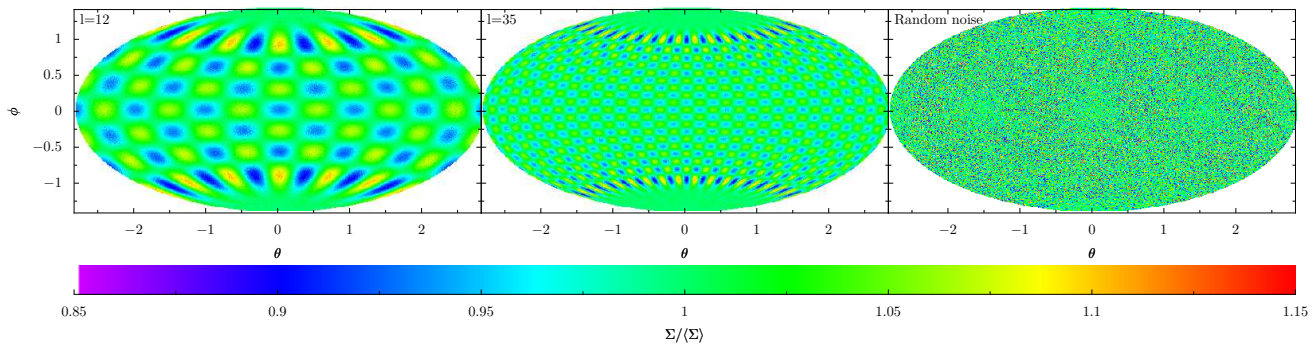}
\caption{Hammer projections of the initial surface density perturbations in the simulation of the evolution of the monochromatic perturbations (left panel: $l=12$, $m=6$, centre panel: $l=35$, $m=17$) and the simulation with random--noise initial conditions (right panel). The colour scale represents fractional deviation of the surface density $\Sigma$ from the mean value $\left<\Sigma\right>$.}
\label{fig:init}
\end{figure*}
\subsection{Non--pressure--confined shells}
\begin{figure*}
     \centering
     \subfigure[Variation with time of $\omega_{12}$ in the SPH (solid blue line) and AMR (dashed blue line) simulations, compared with the analytical value of $\omega_{12}$ (red line).]{\includegraphics[width=0.30\textwidth]{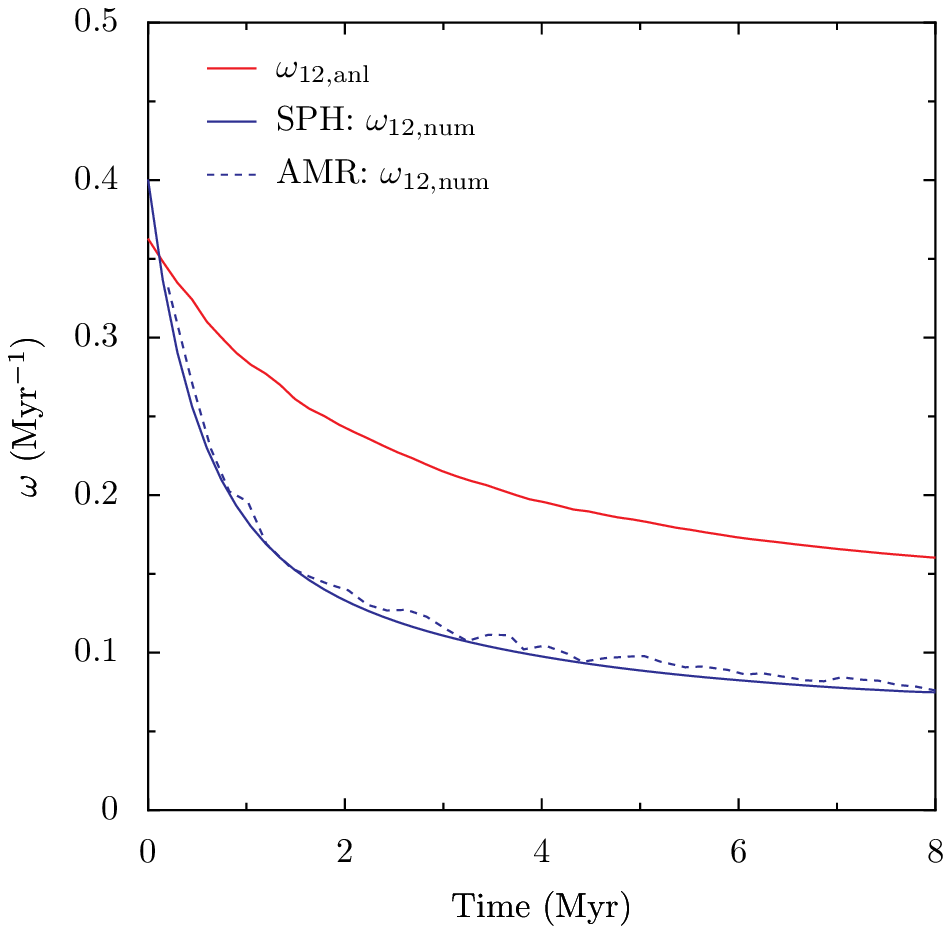}}
\label{fig:l12_omegat_nopress}          
     \hspace{.1in}
     \subfigure[Variation with time of the integrated analytical value of $\omega_{12}(t)$ as derived from Equation \ref{eqn:disp_rel} (magenta line) and the integrated numerical value derived from the SPH (solid blue line) and AMR (dashed blue line) simulations.]{\includegraphics[width=0.30\textwidth]{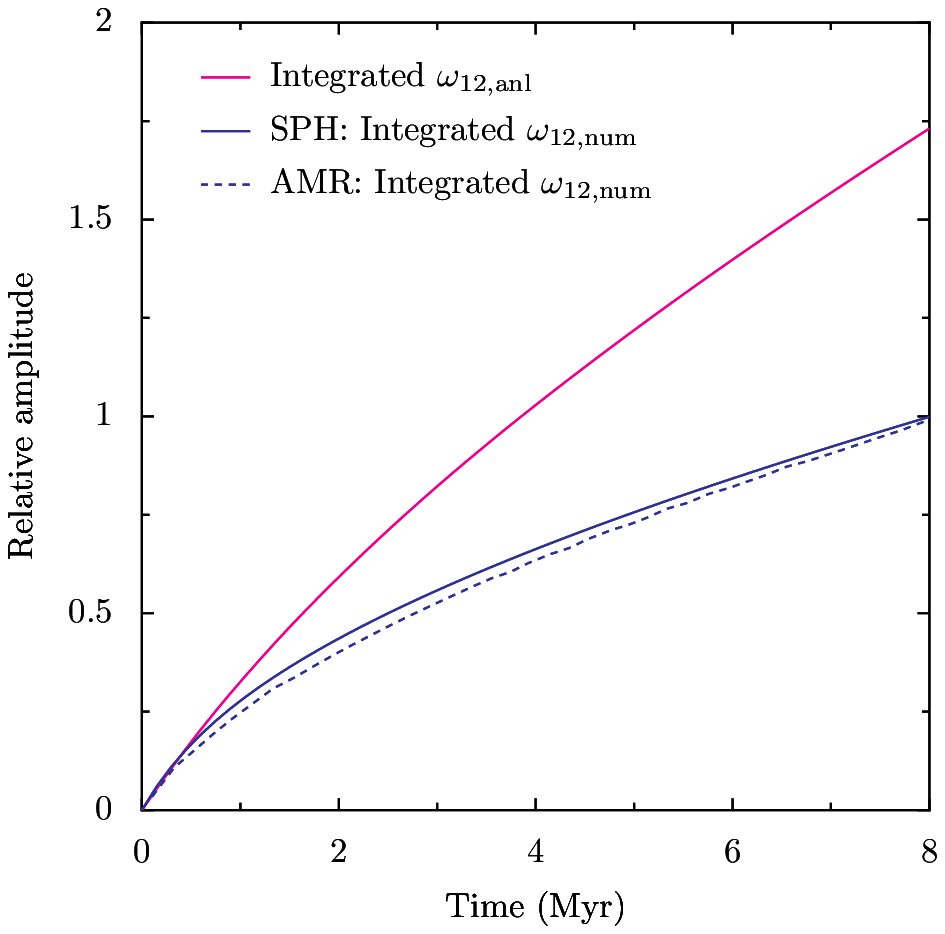}}
\label{fig:l12_omt_nopress}
      \hspace{.1in}
      \subfigure[Variation with time of the mean surface density $\Sigma_{0}$ (cyan line), and the perturbed surface density as computed by $\Sigma_{1}(0)$exp$(I_{\rm f})$ (blue lines) and by $(\Sigma_{\rm max}-\Sigma_{\rm min})/2$ (green lines) from the SPH (solid lines) and AMR (dashed lines) simulations for the $l=12$ perturbation.]{\includegraphics[width=0.30\textwidth]{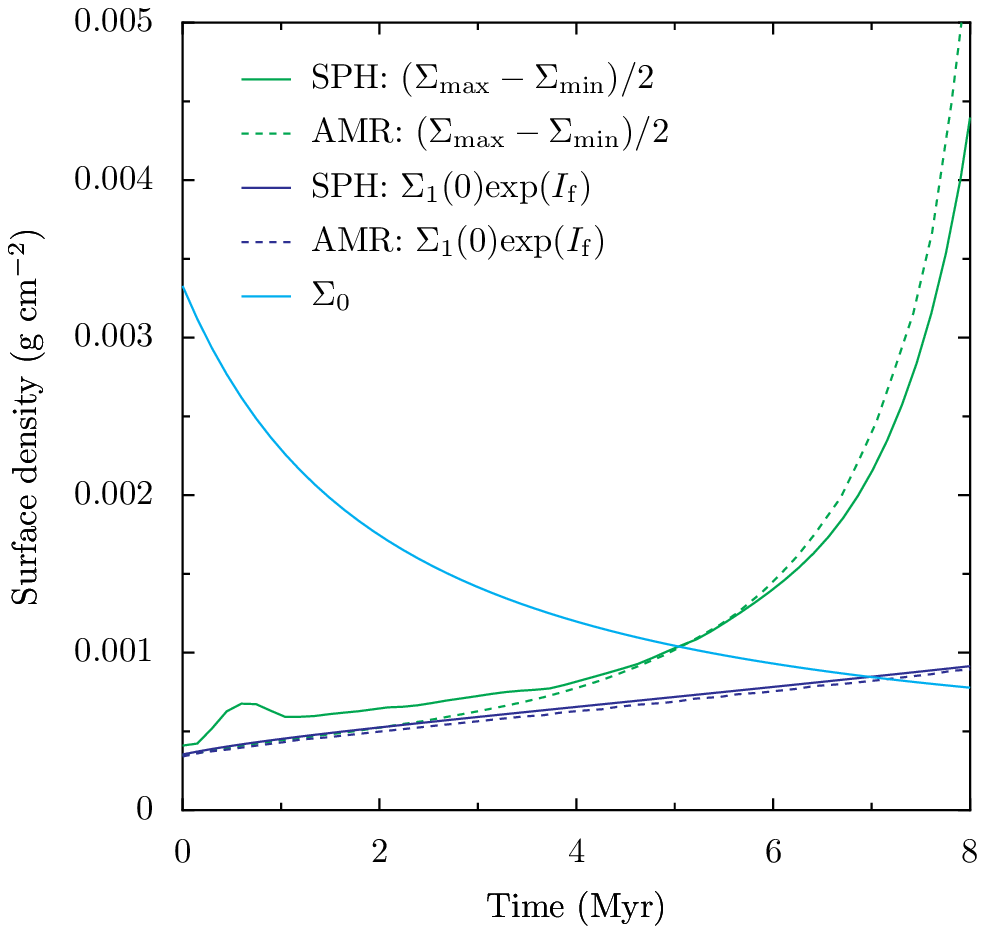}}
\label{fig:l12_sigt_nopress}
     \caption{Evolution of the $l=12$ perturbation in the non--pressure--confined shell.}
     \label{fig:l12_plots}
\end{figure*}     
\begin{figure*}
     \centering
     \subfigure[]{\includegraphics[width=.48\textwidth]{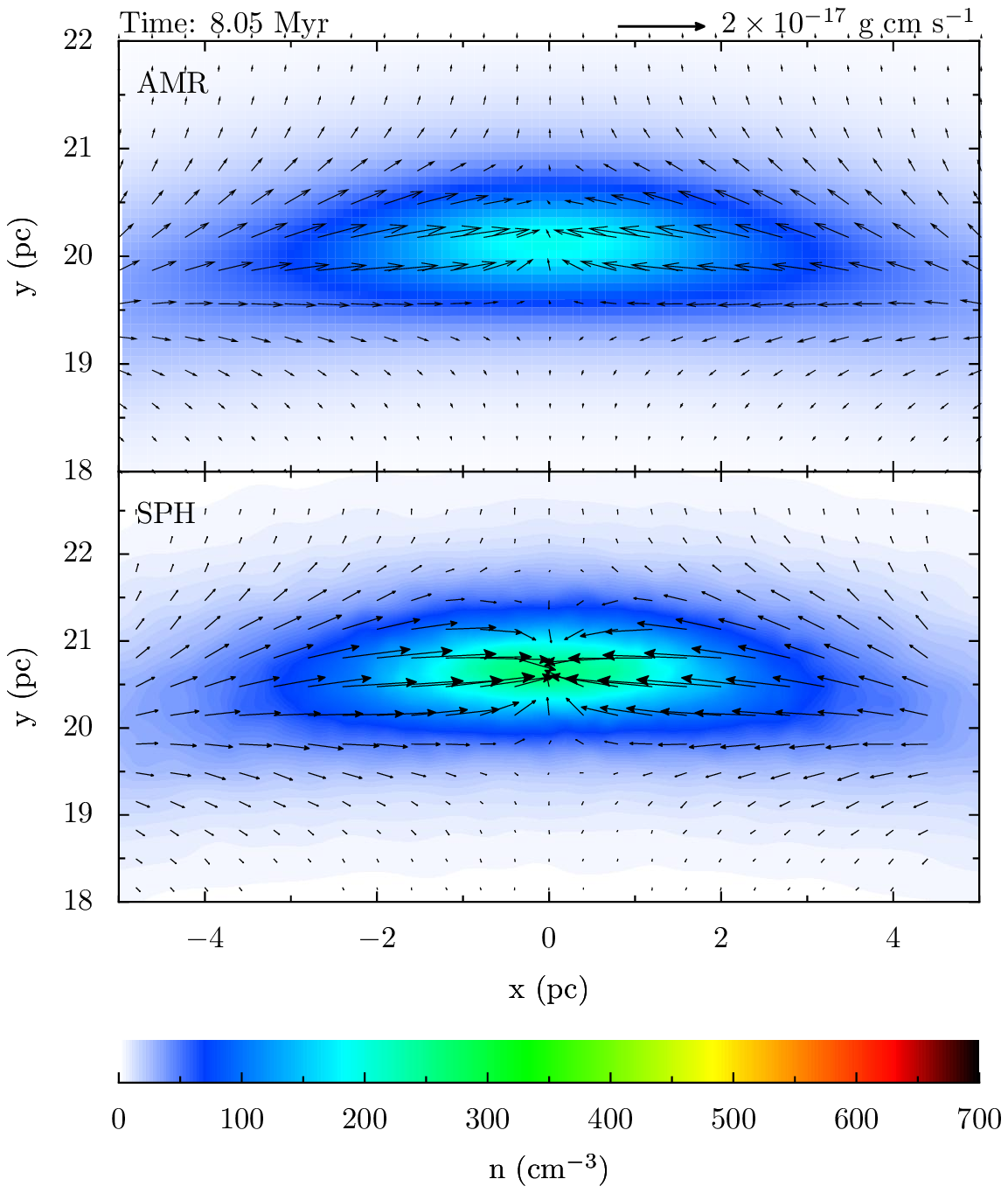}}
     \subfigure[]{\includegraphics[width=.48\textwidth]{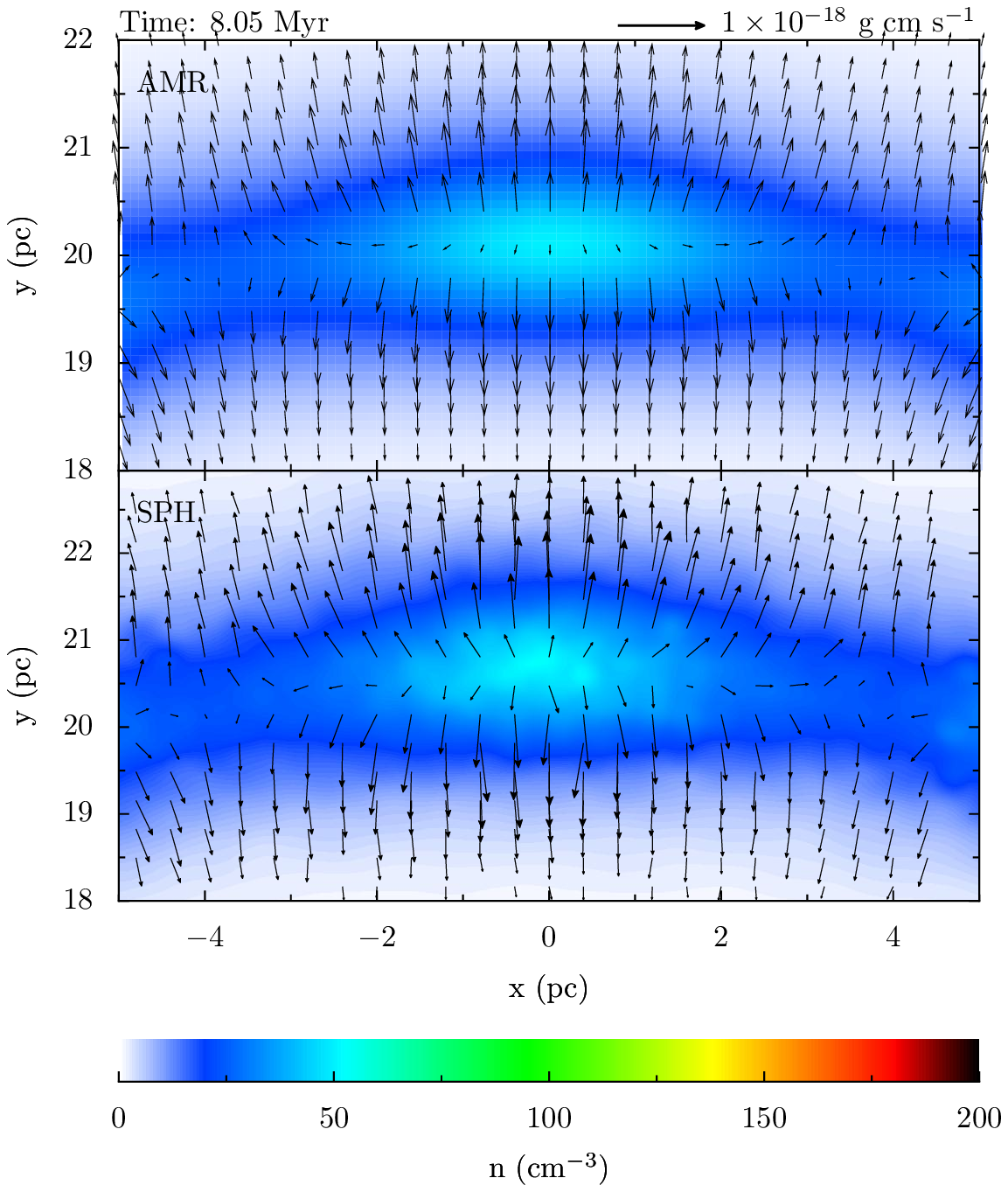}}
     \caption{Slices through the $z=0$ plane of the non--pressure--confined shells perturbed with the $l=12$ (left) and $l=35$ (right) modes at a time of $8$Myr. Top panels are from the AMR calculations, bottom panels are from the SPH runs. Colours represent gas number density, arrows are mass fluxes.}
     \label{fig:lowpress_flows}
\end{figure*} 
\indent The $l=12$, $m=6$ perturbation which we inserted into the shell's surface density is shown in the left panel of Figure \ref{fig:init}. Its evolution is depicted in Figure \ref{fig:l12_plots} and we show a slice through the $z=0$ plane of the shell at a time of $8$Myr in the left half of Figure \ref{fig:lowpress_flows}, showing the gas densities and mass fluxes in the AMR and SPH calculations. The evolution of this mode is similar to, but slower than, that predicted by the thin--shell approximation. In Figures \ref{fig:l12_plots}(a) and \ref{fig:l12_plots}(b), we compare the instantaneous and integrated analytical and numerical growth rates of the $l=12$ mode. As shown in Figure \ref{fig:l12_plots}(c), after $\sim5$Myr the surface density perturbation $\Sigma_{1}=\Sigma_{1}(0)$exp$(I_{\rm f})$, where the fragmentation integral $I_{\rm f}=\int_{0}^{t}\omega(t^{'}){\rm d}t^{'}$, rises to the mean surface density $\Sigma_{0}$ and later the growth of the mode becomes nonlinear. The small oscillation in the surface density perturbation in the SPH calculation at $t\approx0.5$Myr (also visible in the other SPH simulations of monochromatic perturbations) is a result of fluctuations in high--frequency modes as a result of noise in the surface density. The iterative procedure described above cannot completely remove the noise from the SPH calculations without producing an unacceptably large range of particle masses. However, the high degree of consonance between the SPH calculations and the AMR runs, which have a much lower level of intrinsic noise, implies that the noise in the SPH initial conditions has no significant impact on our results.\\
\indent The left half of Figure \ref{fig:lowpress_flows} shows that strong radial flows cause the shell to thicken, but there are also clear azimuthal flows feeding material into the density peaks of the $l=12$ perturbation, allowing it to grow. We see that the density within the fragment is slightly higher in the SPH calculation.\\
\begin{figure*}
     \centering
     \subfigure[Variation with time of the $\omega_{35}$ spherical harmonic multipole coefficient in the SPH (solid blue line) and AMR (dashed blue line) simulations, compared with the analytical value of $\omega_{35}$ (red line).]{\includegraphics[width=0.30\textwidth]{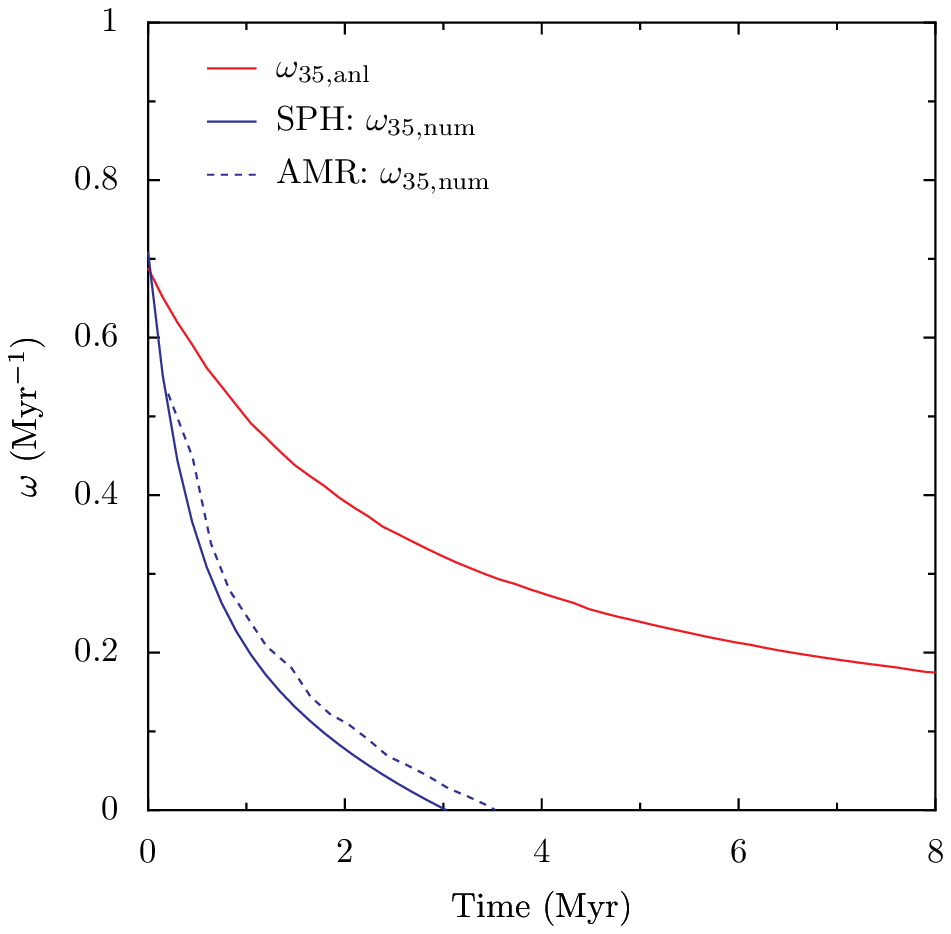}}
\label{fig:l35_clt_nopress}          
     \hspace{.1in}
     \subfigure[Variation with time of the integrated analytical value of $\omega_{35}(t)$ as derived from Equation \ref{eqn:disp_rel} (magenta line) and the integrated numerical value derived from the SPH (solid blue line) and AMR (dashed blue line) simulations.]{\includegraphics[width=0.30\textwidth]{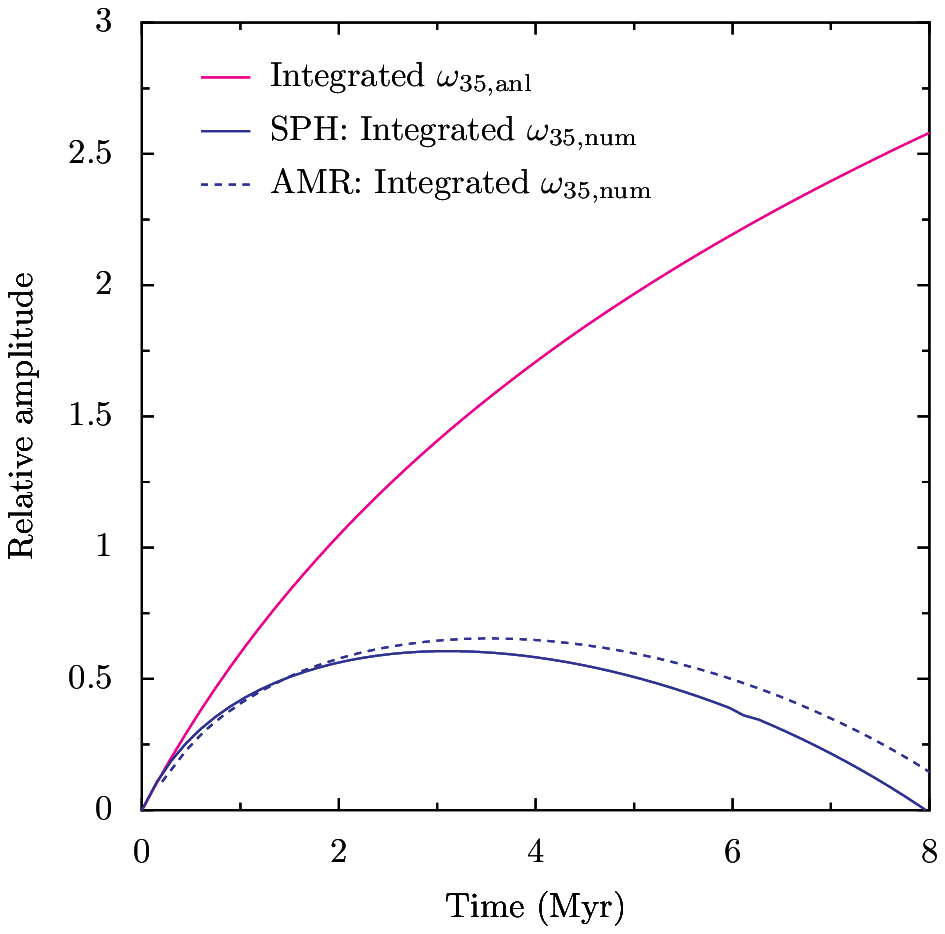}}
\label{fig:l35_omt_nopress}
      \hspace{.1in}
      \subfigure[Variation with time of the mean surface density $\Sigma_{0}$ (cyan line), and the perturbed surface density as computed by $\Sigma_{1}(0)$exp$(I_{\rm f})$ (blue lines) and by $(\Sigma_{\rm max}-\Sigma_{\rm min})/2$ (green lines) from the SPH (solid lines) and AMR (dashed lines) simulations for the $l=35$ perturbation.]{\includegraphics[width=0.30\textwidth]{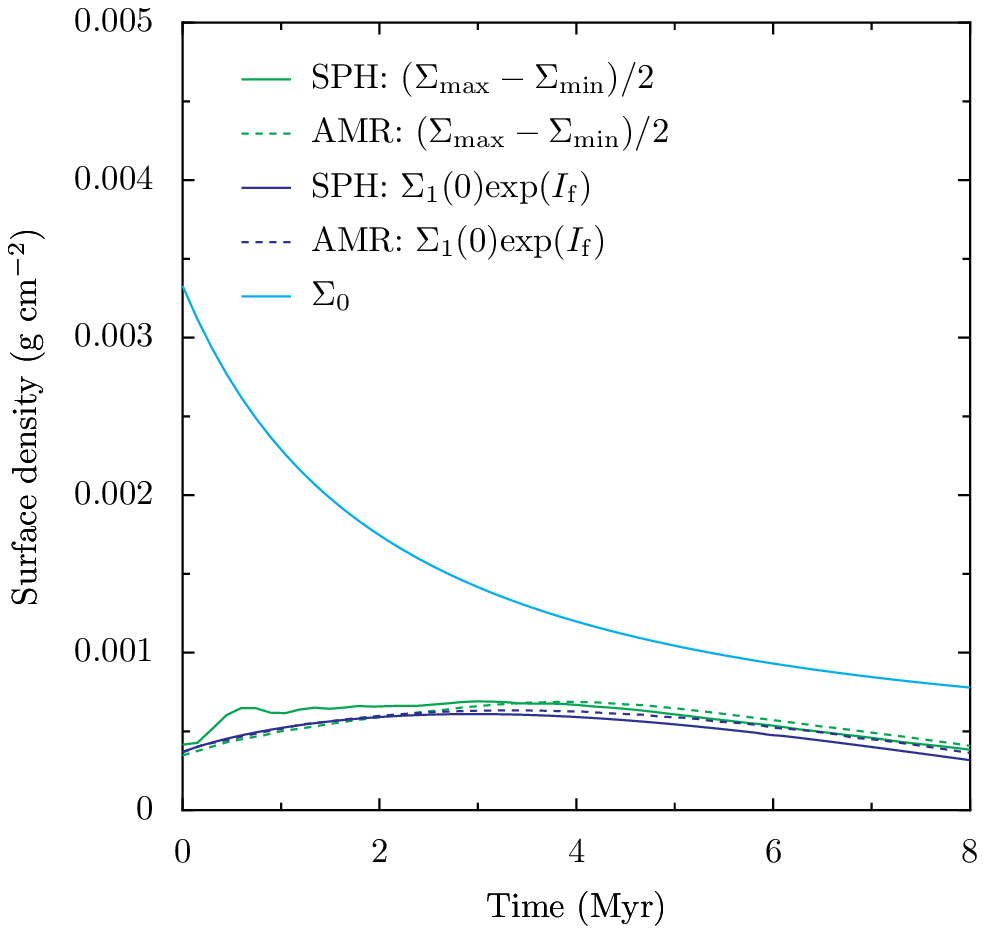}}
\label{fig:l35_sigt_nopress}
     \caption{Evolution of the $l=35$ perturbation in the non--pressure--confined shell.}
     \label{fig:l35_plots}
\end{figure*}
\indent The behaviour of the $l=35$ perturbation, whose initial form is shown in the centre panel of Figure \ref{fig:init} is very different. As shown in Figures \ref{fig:l35_plots}(a) and \ref{fig:l35_plots}(b), although the $l=35$ mode does grow for the first $\sim3$Myr of the calculation, its amplitude then levels off and begins to decrease, dropping below its initial amplitude after $\sim7$Myr.\\
\indent Slices through the $z=0$ plane of the shell, shown in the right half of Figure \ref{fig:lowpress_flows} again show flows perpendicular to the shell, as seen in the simulations of the $l=12$ mode, and these flows, together with weak tangential flows, are transporting material out of the perturbation, causing it to dissolve.\\
\begin{figure*}    
     
     \hspace{0.05in}
   \subfigure[]{\includegraphics[width=.32\textwidth]{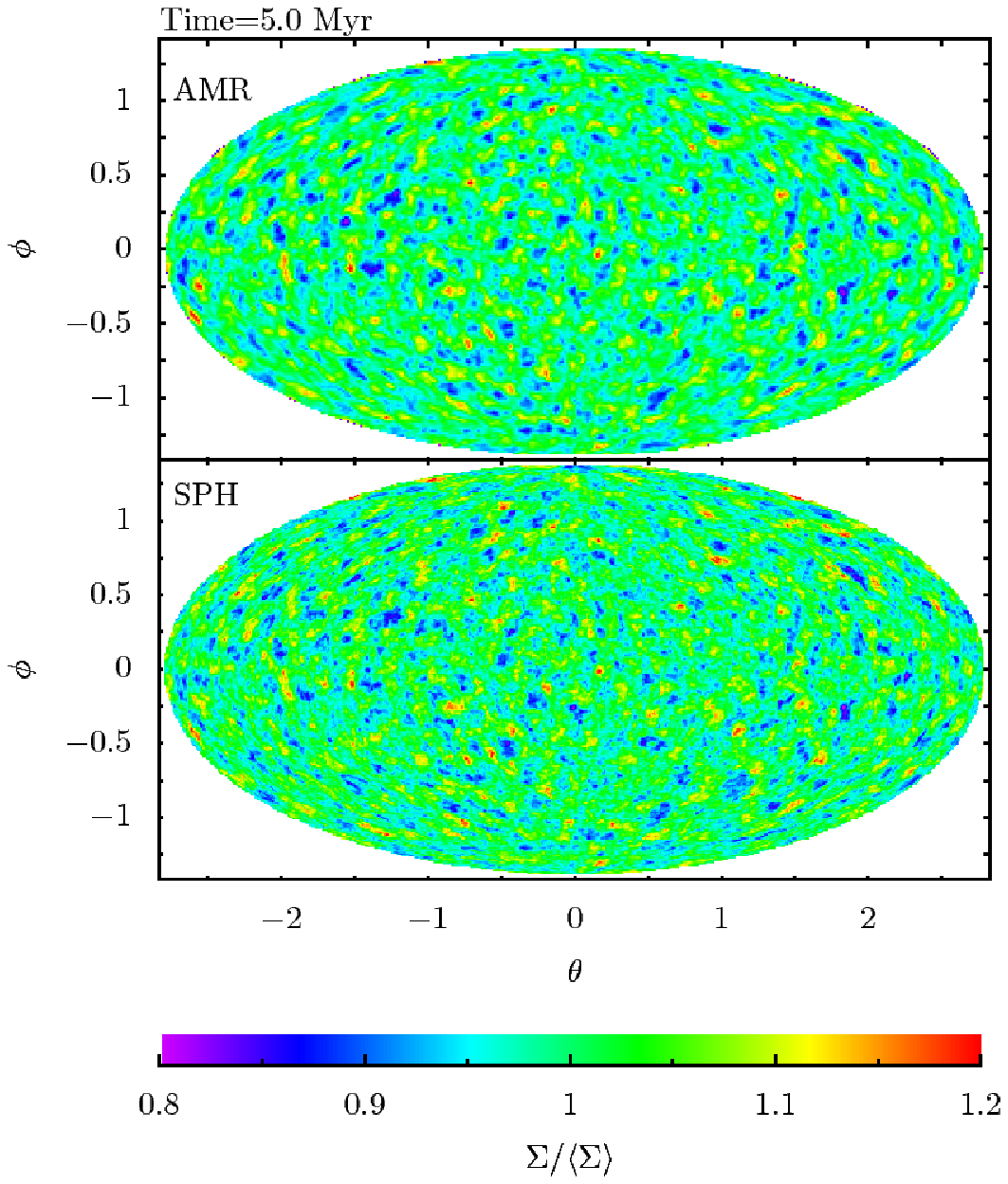}}
     \hspace{0.03in}
    \subfigure[]{\includegraphics[width=.32\textwidth]{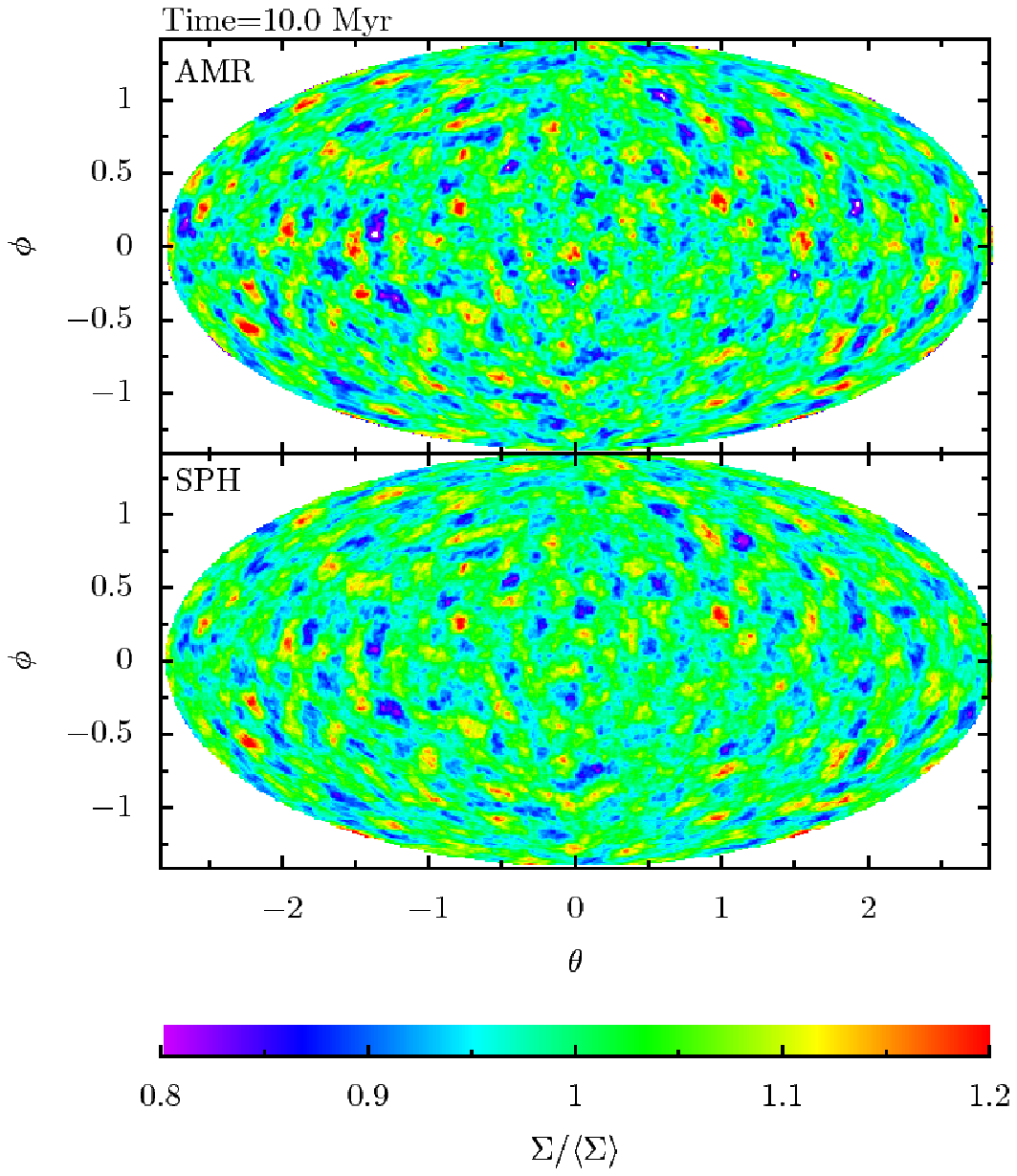}}
     \hspace{0.01in}
     \subfigure[]{\includegraphics[width=.32\textwidth]{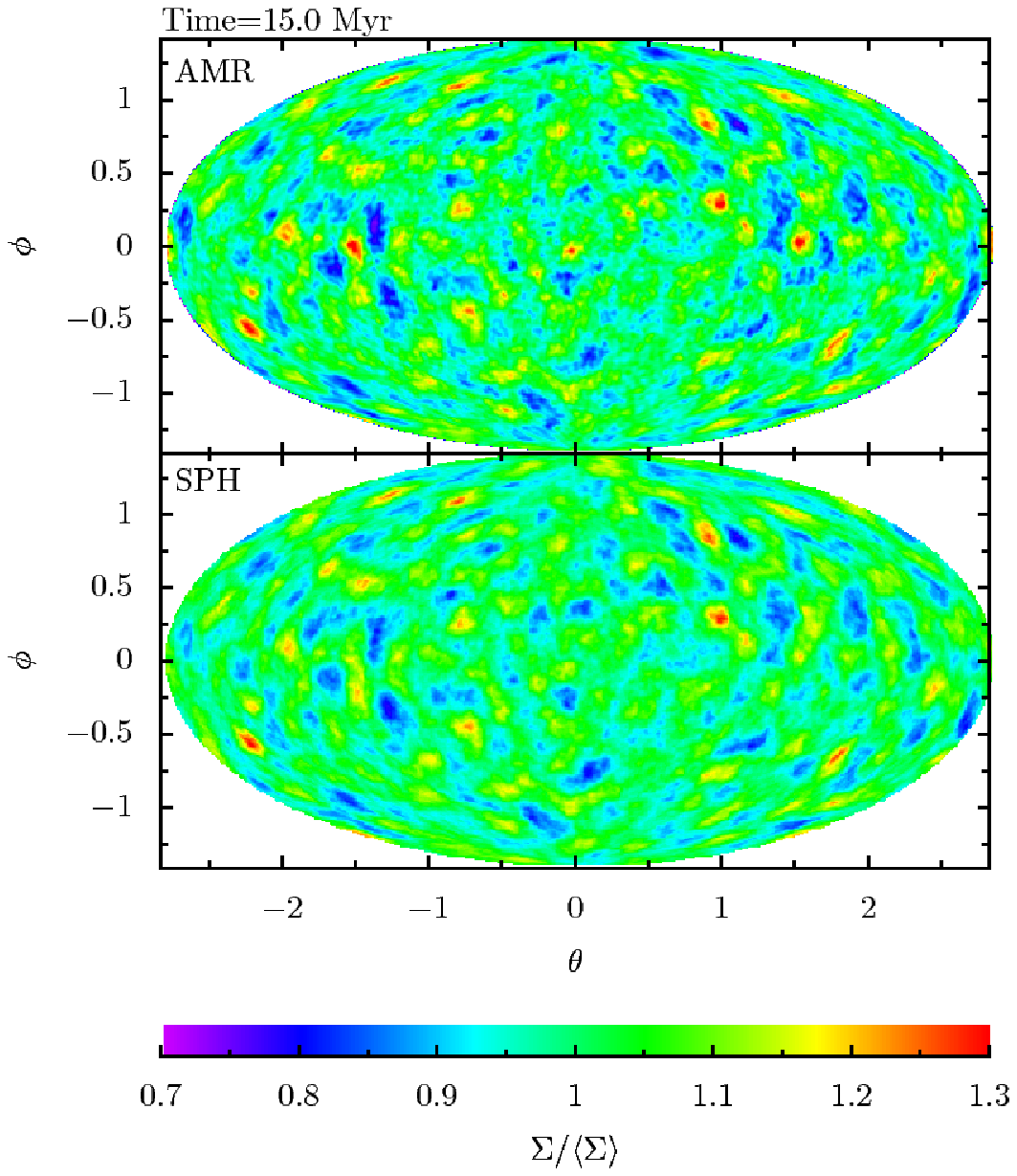}}
     \vspace{0.1in}
     \subfigure[]{\includegraphics[width=.33\textwidth]{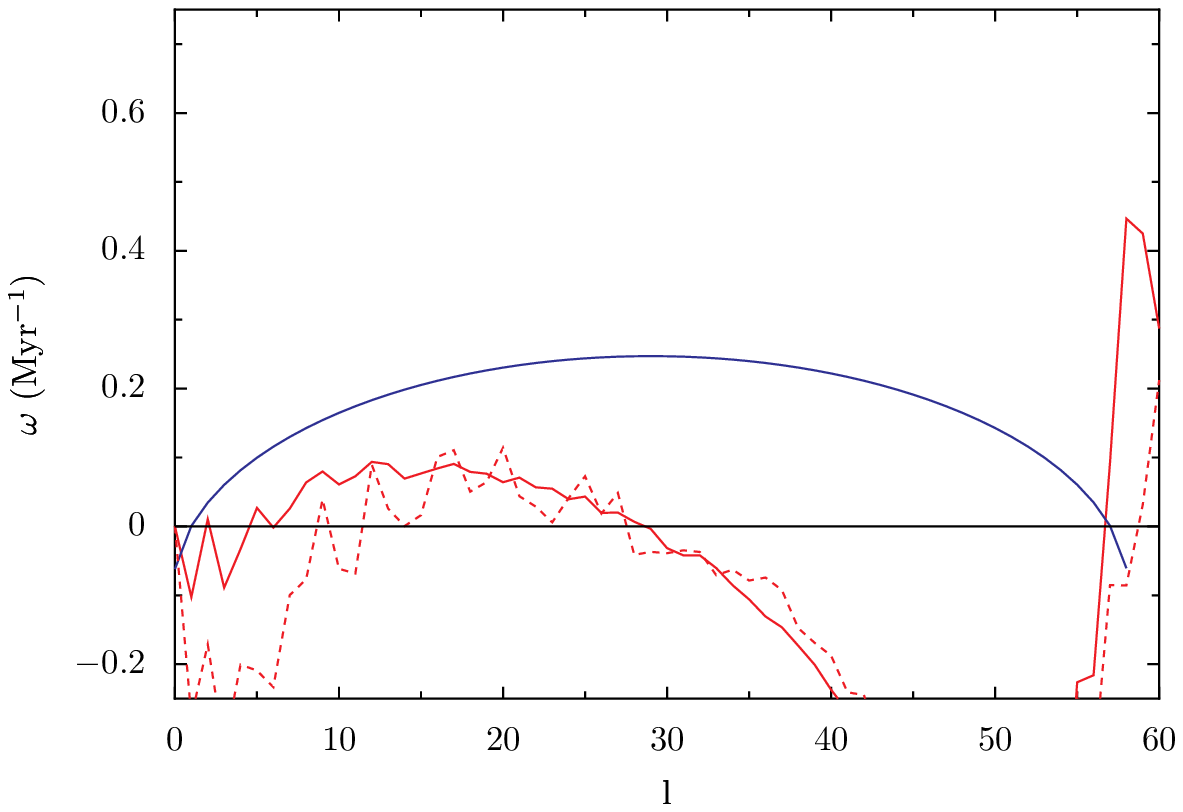}}
     \subfigure[]{\includegraphics[width=.33\textwidth]{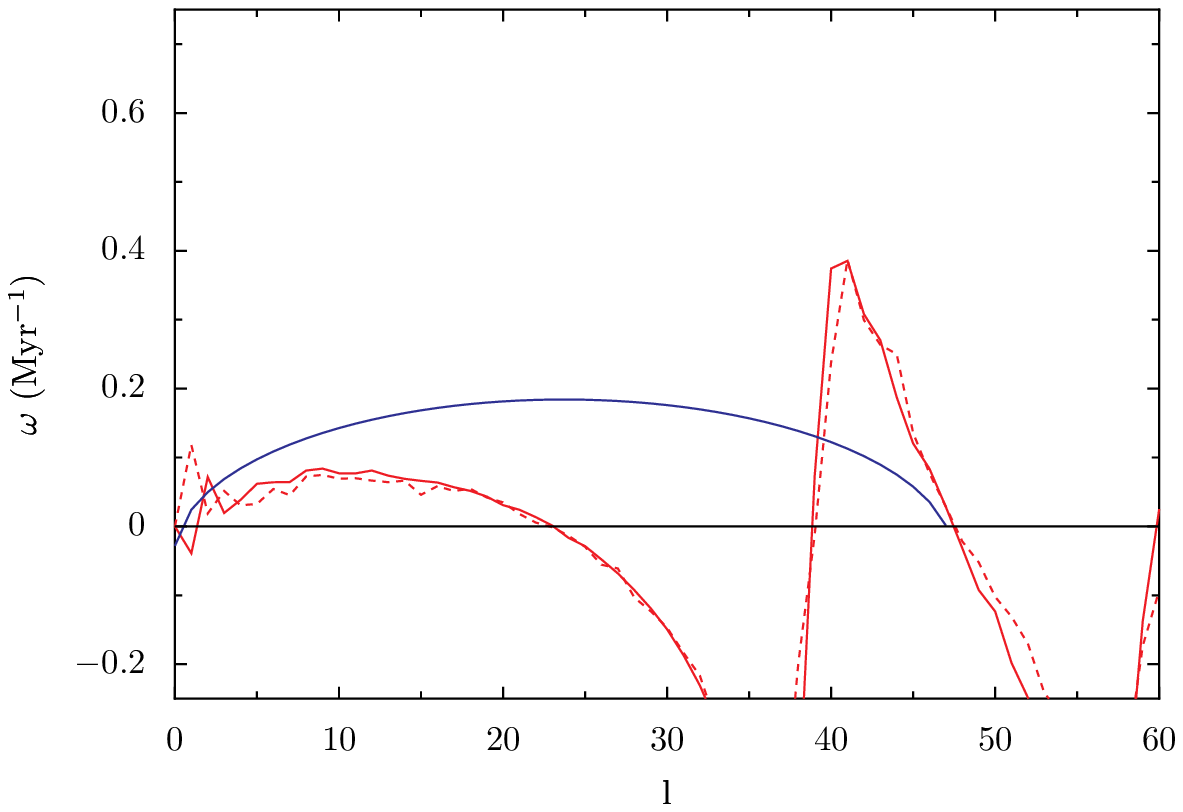}}
     \subfigure[]{\includegraphics[width=.33\textwidth]{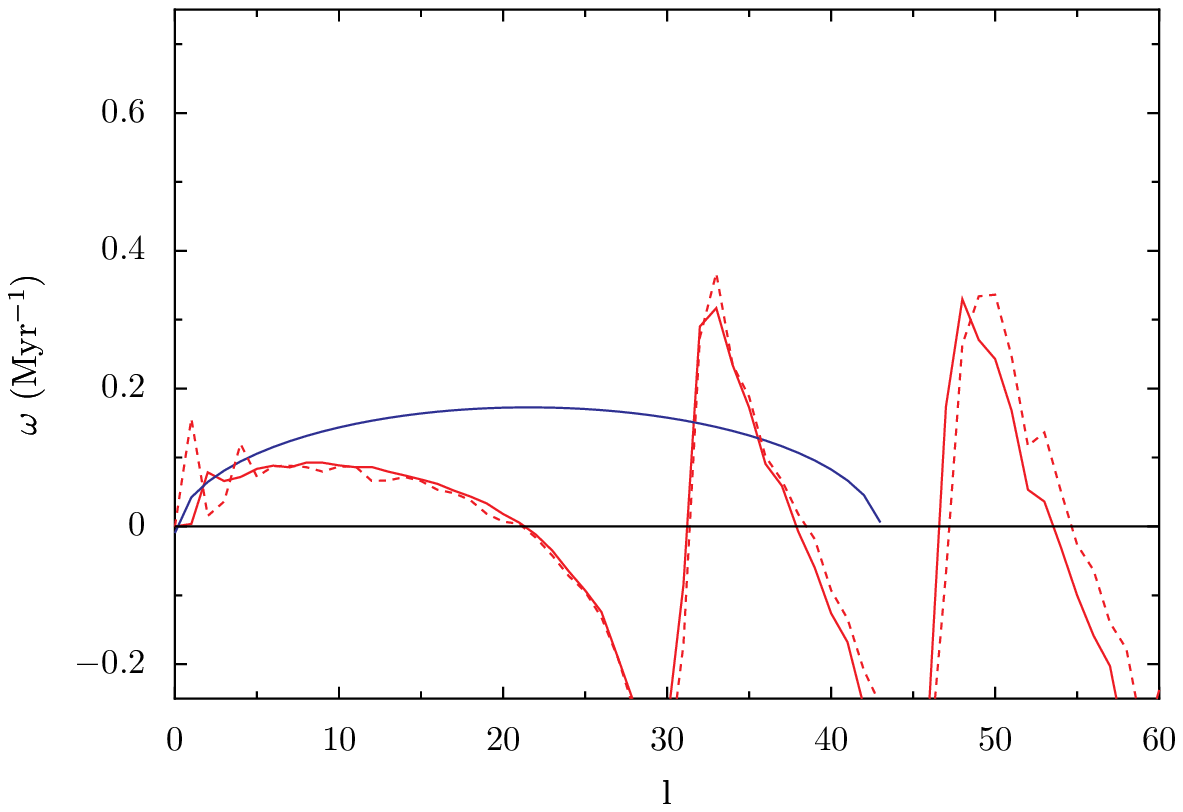}}

\caption{Hammer plots from the SPH and AMR simulations of the non--pressure--confined shell seeded with a Maxwellian random velocity field at three epochs (a--c) and comparisons of $\omega(l)$ derived from the SPH calculation (red solid line) and from the AMR calculation (red dotted line) with the analytical dispersion relation (solid blue line) at the same epochs (d--f). Only the real part of $\omega$ is shown.}
     \label{fig:noise_nopress}
\end{figure*}
\indent The right panel of Figure \ref{fig:init} shows the initial surface density perturbations in the shell with random noise in the initial density field. Hammer projections of the surface density perturbations and dispersion relations at three epochs from the SPH and FLASH calculations are shown in Figure \ref{fig:noise_nopress}. The codes agree very well with each other. Many features in the surface density perturbation are common to the output from both codes and there is also good agreement in the forms of the numerical dispersion relations, although there is some discrepancy in the growth rate of long--wavelength modes at early times. We observe that filamentary structures form on the surface of the shell which further fragment into strings of approximately circular objects. In both calculations, fragmentation at all wavelengths is slower than predicted by the thin--shell model and fragmentation at $l>20$ is suppressed entirely. Note that the sharper peaks at higher $l$--values are higher harmonics due to the non--Gaussianity of the forming fragments at later times, and do not reflect structure forming on these scales. These results are in accord with those of our calculations involving monochromatic perturbations, which showed that low--frequency modes grow, although slower than expected, but that the high--frequency modes ($l\geq20$) do not grow at all.\\
\subsection{Pressure--confined shells}
The thickening of the shell caused by its expansion due to the overpressure of the shell gas expanding into the surrounding vacuum and by the decreasing surface density seems to influence the progress of fragmentation. In our second set of simulations, we study the evolution of shells perturbed in the same way as in the previous section, but pressure--confined such that the shell thickness remains approximately constant during the evolution.\\
\begin{figure*}
     \centering
     \subfigure[Variation with time of $\omega_{12}$ in the SPH (solid blue line) and AMR (dashed blue line) simulations, compared with the analytical value of $\omega_{12}$ (red line).]{\includegraphics[width=0.30\textwidth]{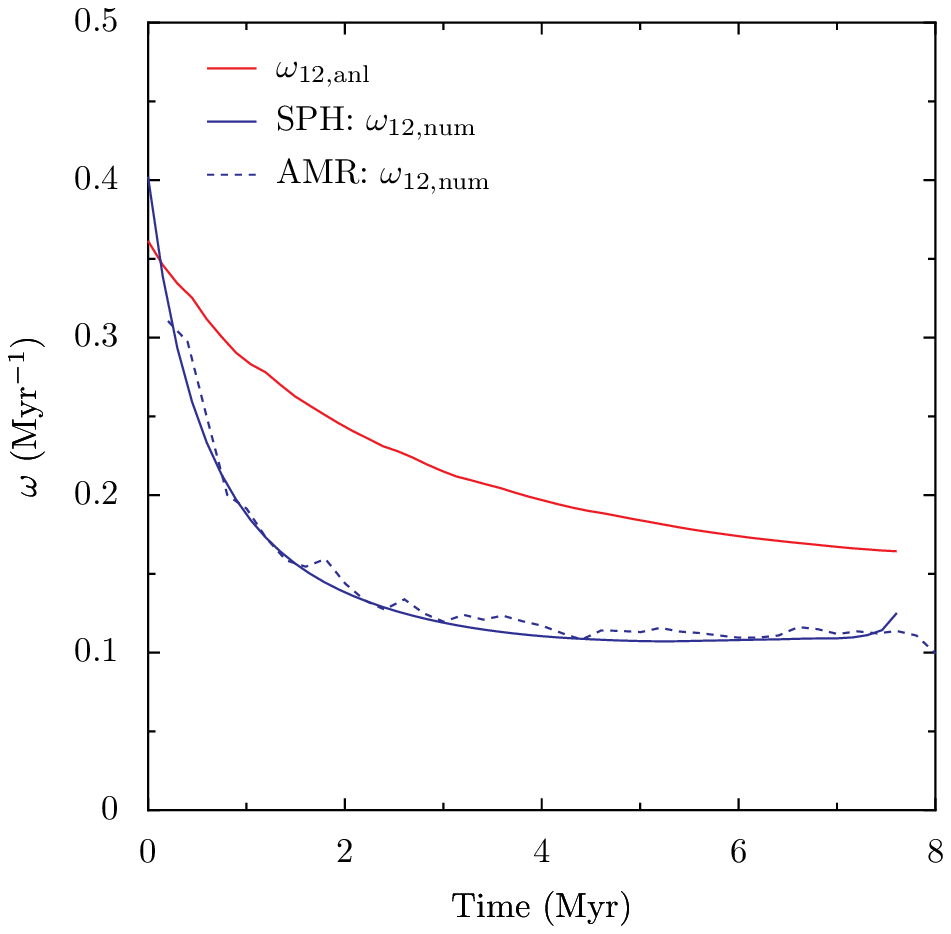}}
\label{fig:l12_clt_press}          
     \hspace{.1in}
     \subfigure[Variation with time of the integrated analytical value of $\omega_{12}(t)$ as derived from Equation \ref{eqn:disp_rel} (magenta line) and the integrated numerical value derived from the SPH (solid blue line) and AMR (dashed blue line) simulations.]{\includegraphics[width=0.30\textwidth]{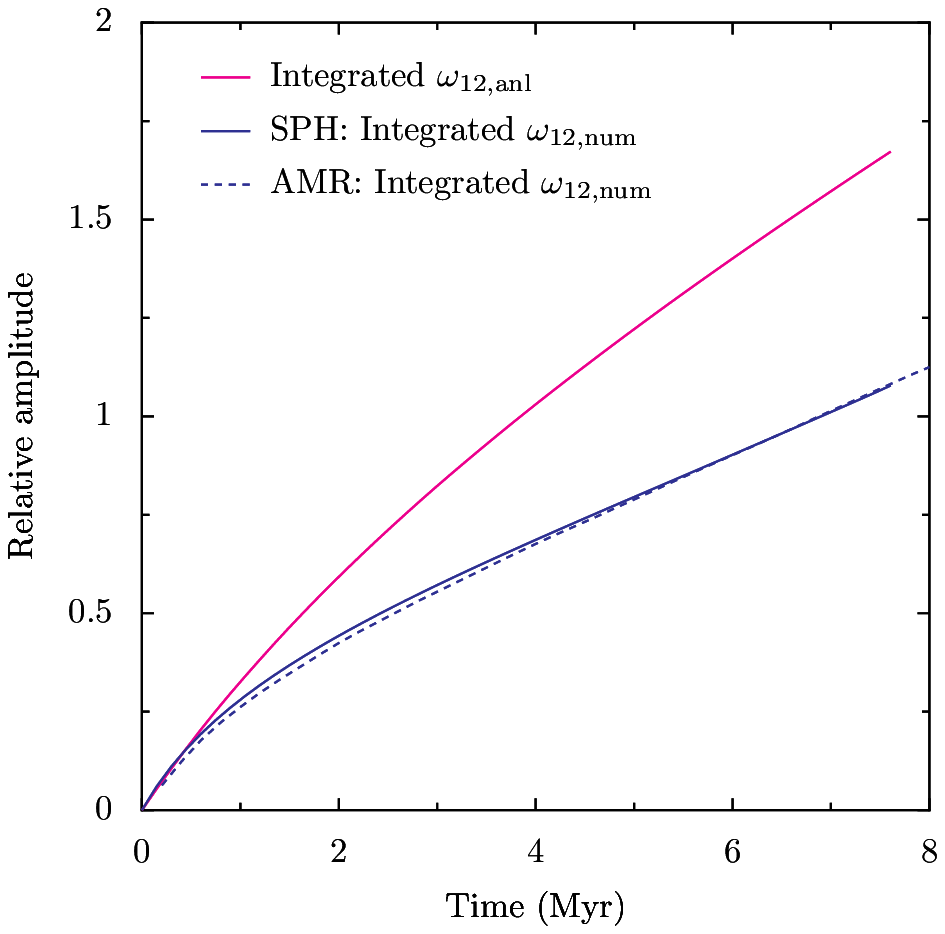}}
\label{fig:l12_omt_press}
      \hspace{.1in}
      \subfigure[Variation with time of the mean surface density $\Sigma_{0}$ (cyan line), and the perturbed surface density as computed by $\Sigma_{1}(0)$exp$(I_{\rm f})$ (blue lines) and by $(\Sigma_{\rm max}-\Sigma_{\rm min})/2$ (green lines) from the SPH (solid lines) and AMR (dashed lines) simulations for the $l=12$ perturbation.]{\includegraphics[width=0.30\textwidth]{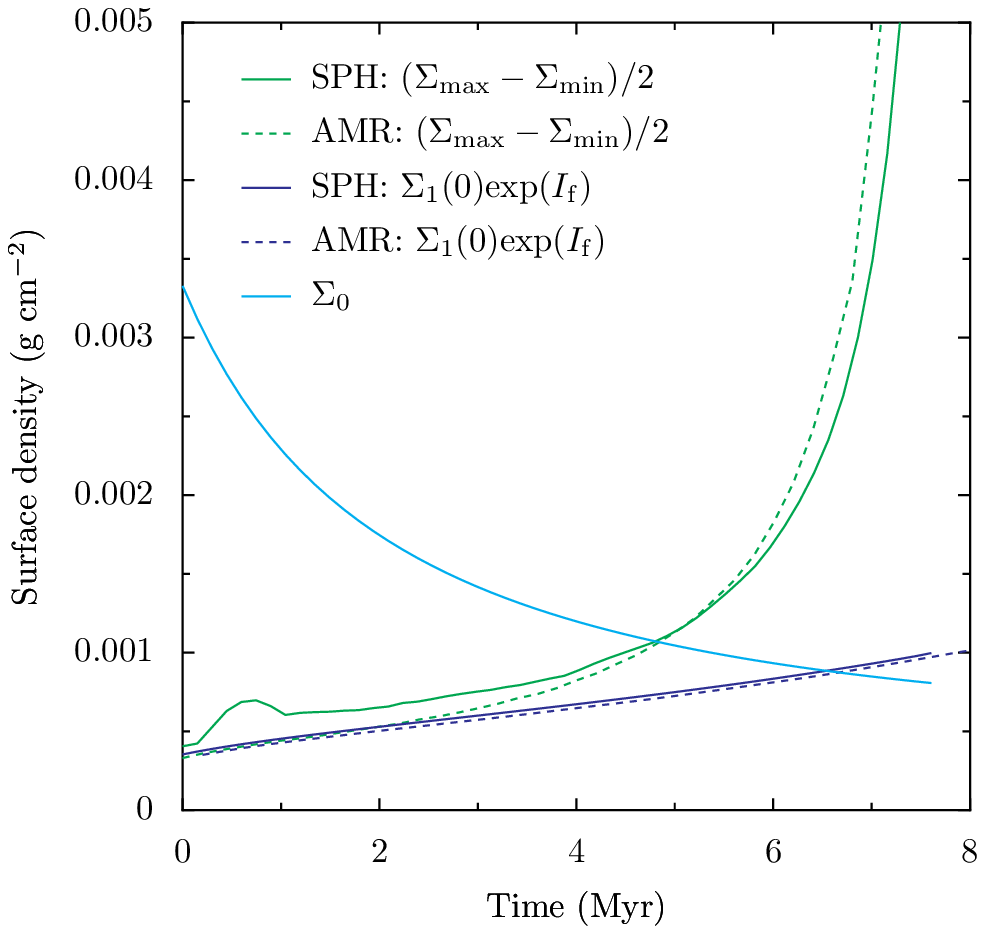}}
\label{fig:l12_sigt_press}
     \caption{Evolution of the $l=12$ perturbation in the pressure--confined shell.}
     \label{fig:l12_press_plots}
\end{figure*}     
\begin{figure*}
     \centering
     \subfigure[]{\includegraphics[width=.48\textwidth]{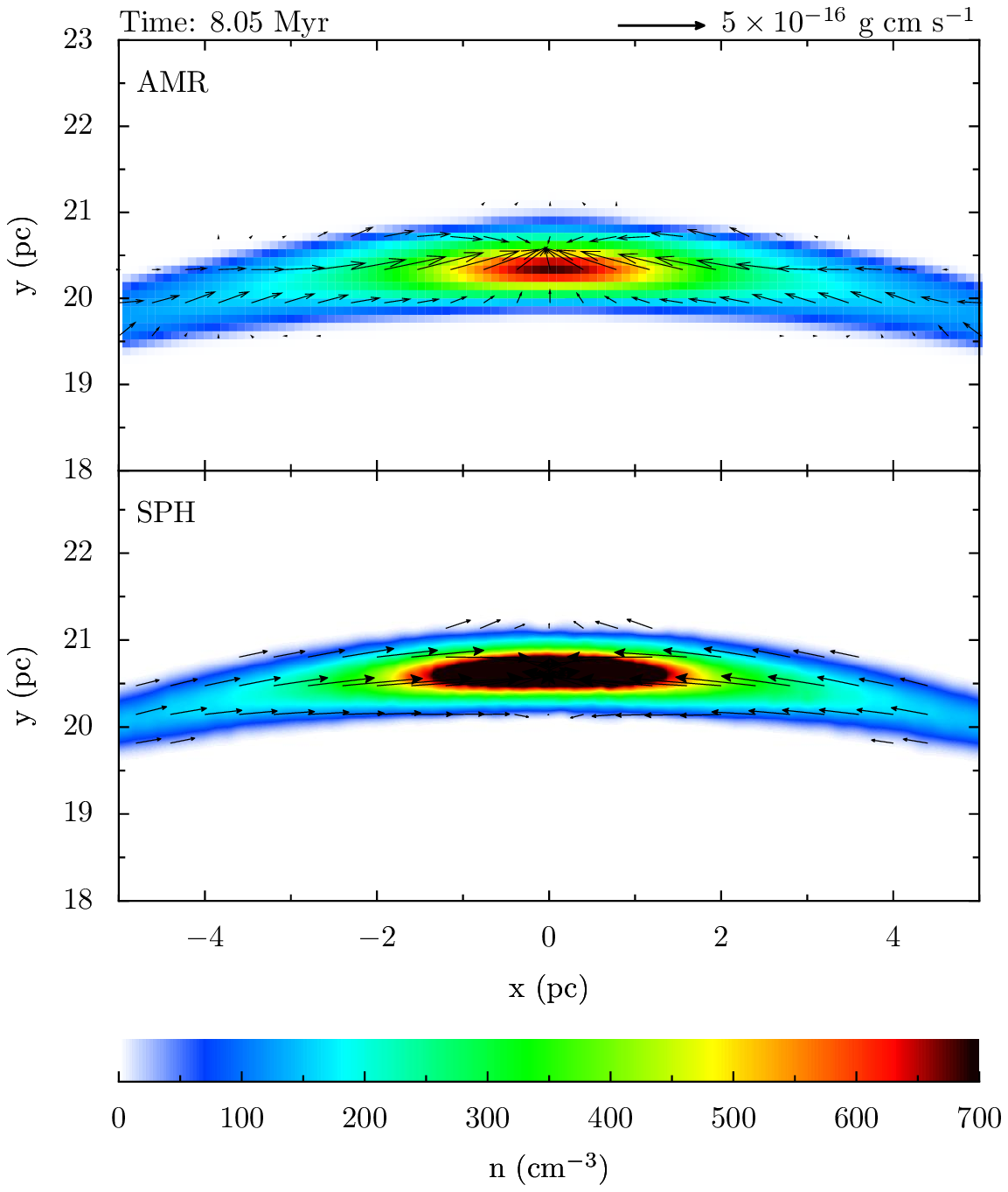}}
     \subfigure[]{\includegraphics[width=.48\textwidth]{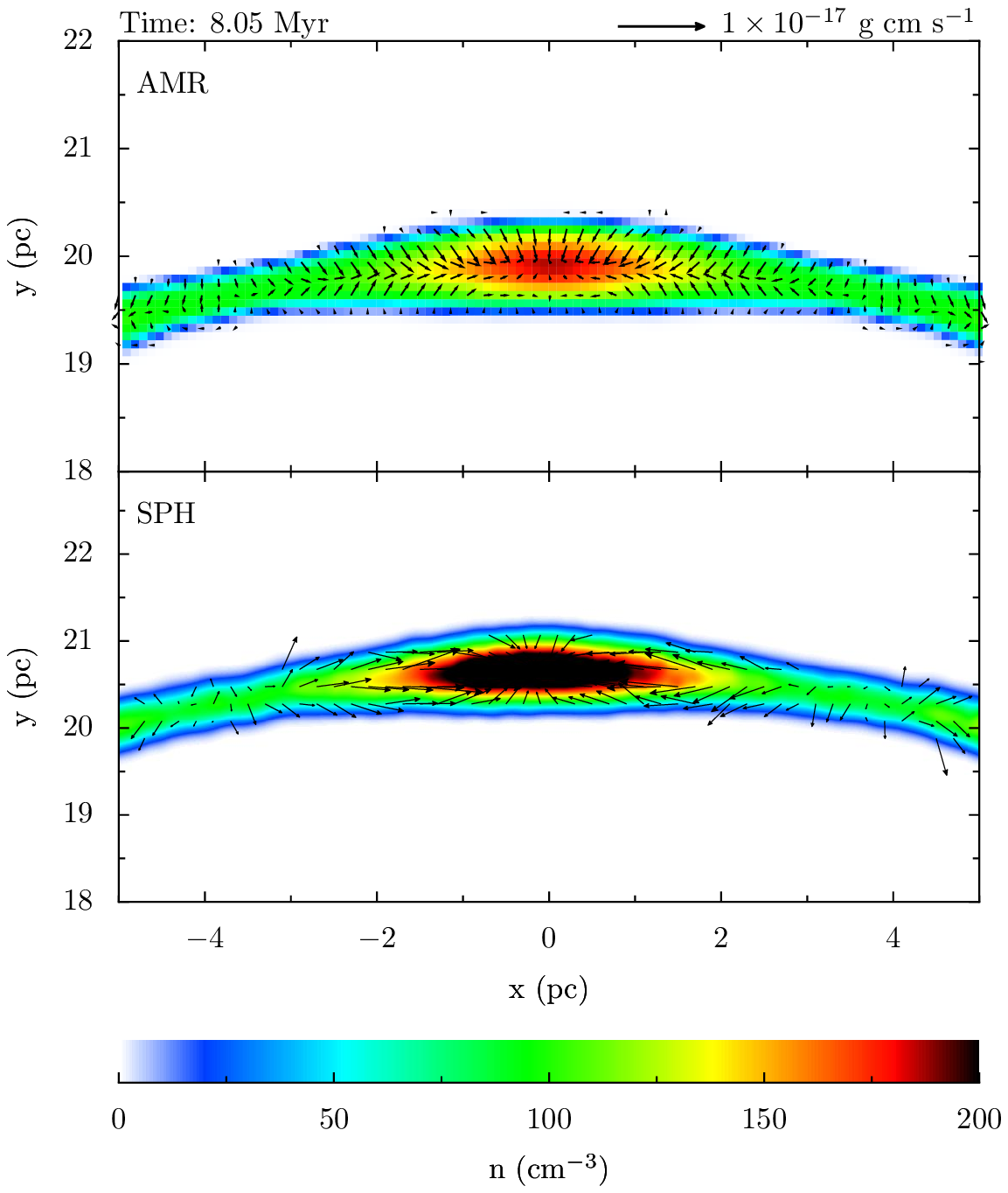}}
     \caption{Slices through the $z=0$ plane of the pressure--confined shells perturbed with the $l=12$ (left) and $l=35$ (right) modes at a time of $8$Myr. Top panels are from the AMR calculations, bottom panels are from the SPH runs. Colours represent gas number density, arrows are mass fluxes.}
     \label{fig:press_flows}
\end{figure*}
\indent The evolution of the pressure--confined shells with the $l=12,m=6$ perturbations is very similar to that of the same simulations without pressure confinement. The AMR and SPH simulations are in good agreement and once again show that the growth of the mode is slower than predicted by the thin--shell analysis. The evolution of the surface density perturbation (Figures \ref{fig:l12_press_plots}(c)) is also very similar between the pressure--confined and non--pressure--confined simulations. Slices through a section of the $z=0$ plane show that the confining pressure largely prevents the thickening of the shell in the radial direction seen in the earlier simulations; azimuthal flows are visible at all stages of the simulations. In the left half of Figure \ref{fig:press_flows}, we observe that, instead of the radial flows transporting material away from the middle of the shell, these flows are directed towards the middle of the shell, increasing the peak density and helping to feed material into the density perturbations. This behaviour is due to the constant external pressure coming to dominate the decreasing pressure inside the shell, as predicted in Figure \ref{fig:z0}. With the chosen external pressure of $10^{-13}$ dyne cm$^{-3}$, the shell initially thickens slightly, but then decreases in thickness from a time of $\sim2.5$Myr.\\
\indent By contrast, the $l=35,m=17$ perturbation behaves very differently in the pressure--confined shell than in the shell with negligible confining pressure. Although initially, the growth of this mode is much slower than predicted by the thin--shell approximation, the growth rate eventually accelerates, at around the time when the shell reaches its maximum thickness and begins to get thinner. As with the longer--wavelength perturbation, the growth of the $l=35$ mode eventually leads to nonlinear behaviour when the surface density perturbation due to the mode comes to exceed the mean shell surface density.\\
\indent The reason for the markedly different behaviour of this mode in the pressure--confined shell is shown in the right half of Figure \ref{fig:press_flows}. We see that, contrary to the simulation of the non--pressure--confined shell, azimuthal flows and radial flows towards the middle of the shell feed material into the density peaks of the $l=35$ perturbation, allowing them to grow. We again observe that the peak density is higher in the SPH simulation, which is partly due to the SPH shell being slightly thinner than the AMR shell, due to the different ways in which the external pressure is treated in the two codes.\\
\begin{figure*}
     \centering
     \subfigure[Variation with time of $\omega_{35}$ in the SPH (solid line) and AMR (dashed line) simulations, compared with the analytical value of $\omega_{35}$ (red line).]{\includegraphics[width=0.30\textwidth]{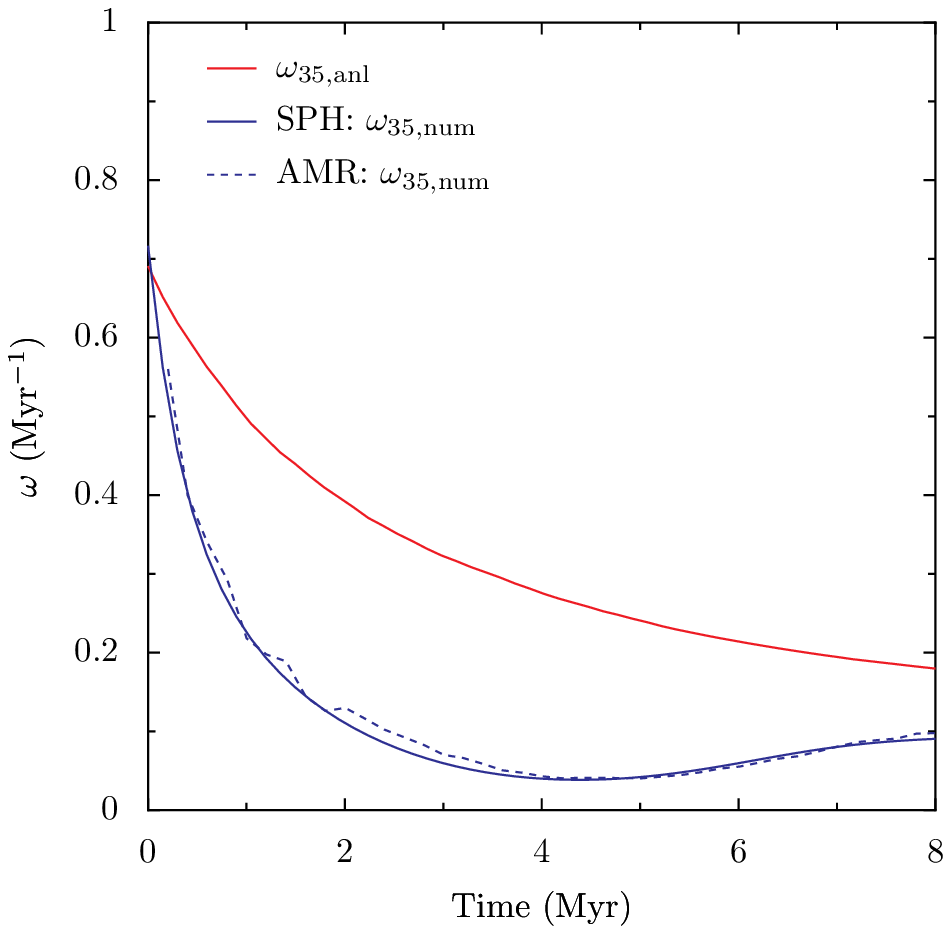}}
\label{fig:l35_clt_press}          
     \hspace{.1in}
     \subfigure[Variation with time of the integrated analytical value of $\omega_{35}(t)$ as derived from Equation \ref{eqn:disp_rel} (magenta line) and the integrated numerical value derived from the SPH (solid blue line) and AMR (dashed blue line) simulations.]{\includegraphics[width=0.30\textwidth]{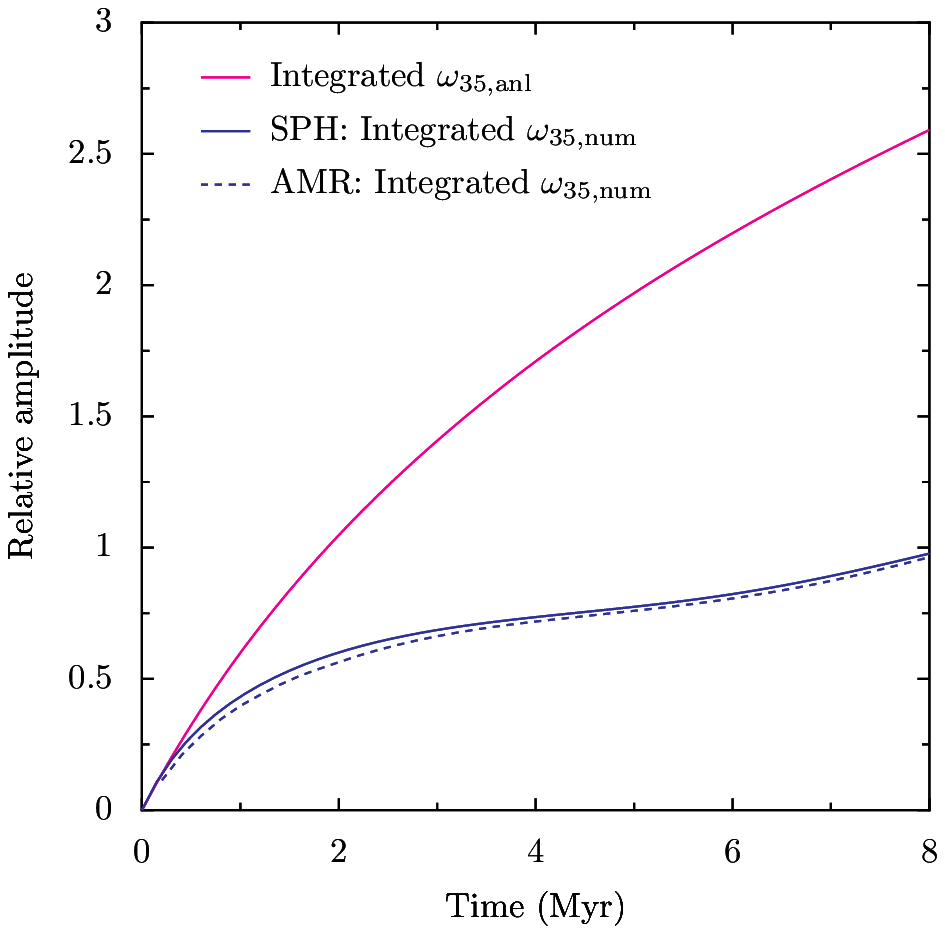}}
\label{fig:l35_omt_press}
      \hspace{.1in}
      \subfigure[Variation with time of the mean surface density $\Sigma_{0}$ (cyan line), and the perturbed surface density as computed by $\Sigma_{1}(0)$exp$(I_{\rm f})$ (blue lines) and by $(\Sigma_{\rm max}-\Sigma_{\rm min})/2$ (green lines) from the SPH (solid lines) and AMR (dashed lines) simulations for the $l=35$ perturbation.]{\includegraphics[width=0.30\textwidth]{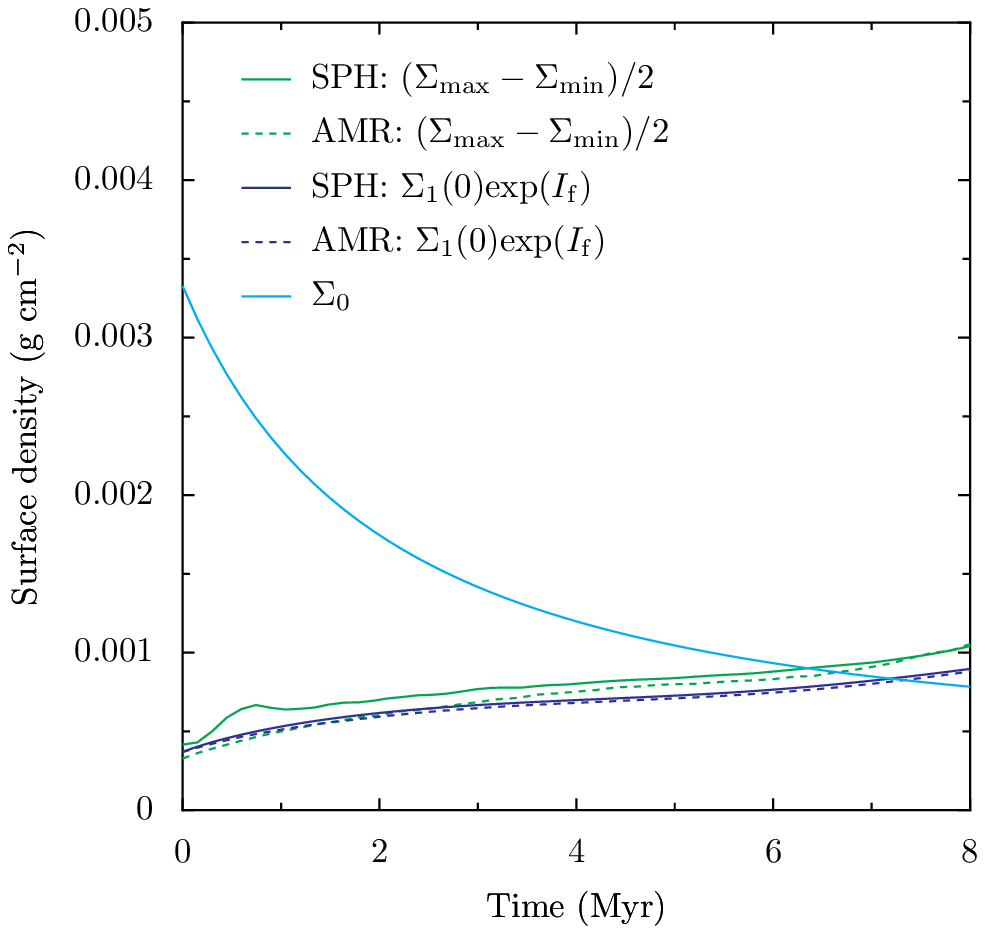}}
\label{fig:l35_sigt_press}
     \caption{Evolution of the $l=35$ perturbation in the pressure--confined shell.}
     \label{fig:l35_press_plots}
\end{figure*}     
\indent The results of the two previous simulations imply that fragmentation in the pressure--confined shell proceeds in a manner more consonant with the predictions of the thin--shell approximation. To see if this impression is correct, we turn to the simulations of the shell with the white noise in the initial density field and the Maxwellian velocity distribution.\\
\begin{figure*}    
     
     \hspace{0.05in}
   \subfigure[]{\includegraphics[width=.32\textwidth]{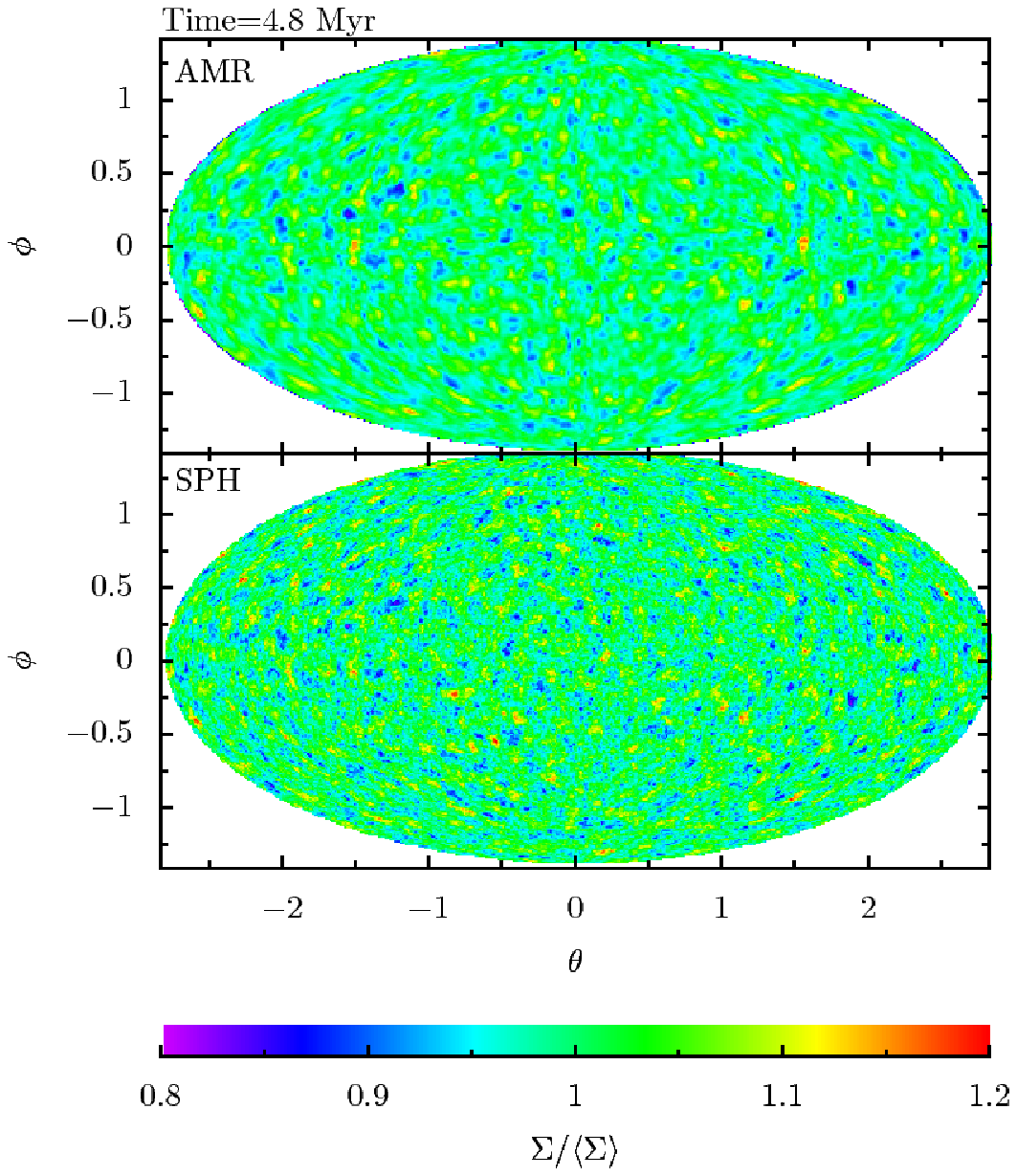}}
     \hspace{0.03in}
    \subfigure[]{\includegraphics[width=.32\textwidth]{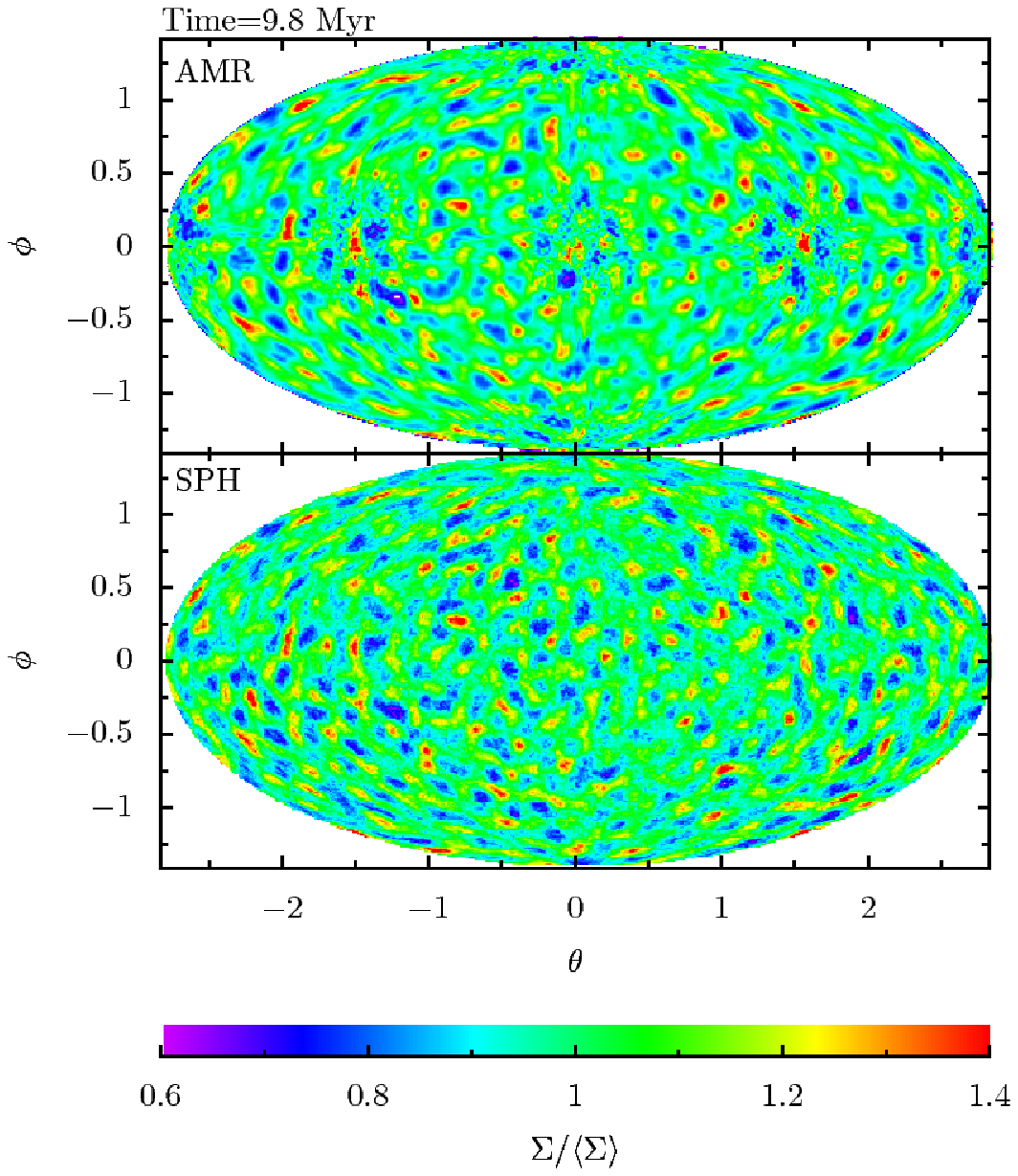}}
     \hspace{0.01in}
     \subfigure[]{\includegraphics[width=.32\textwidth]{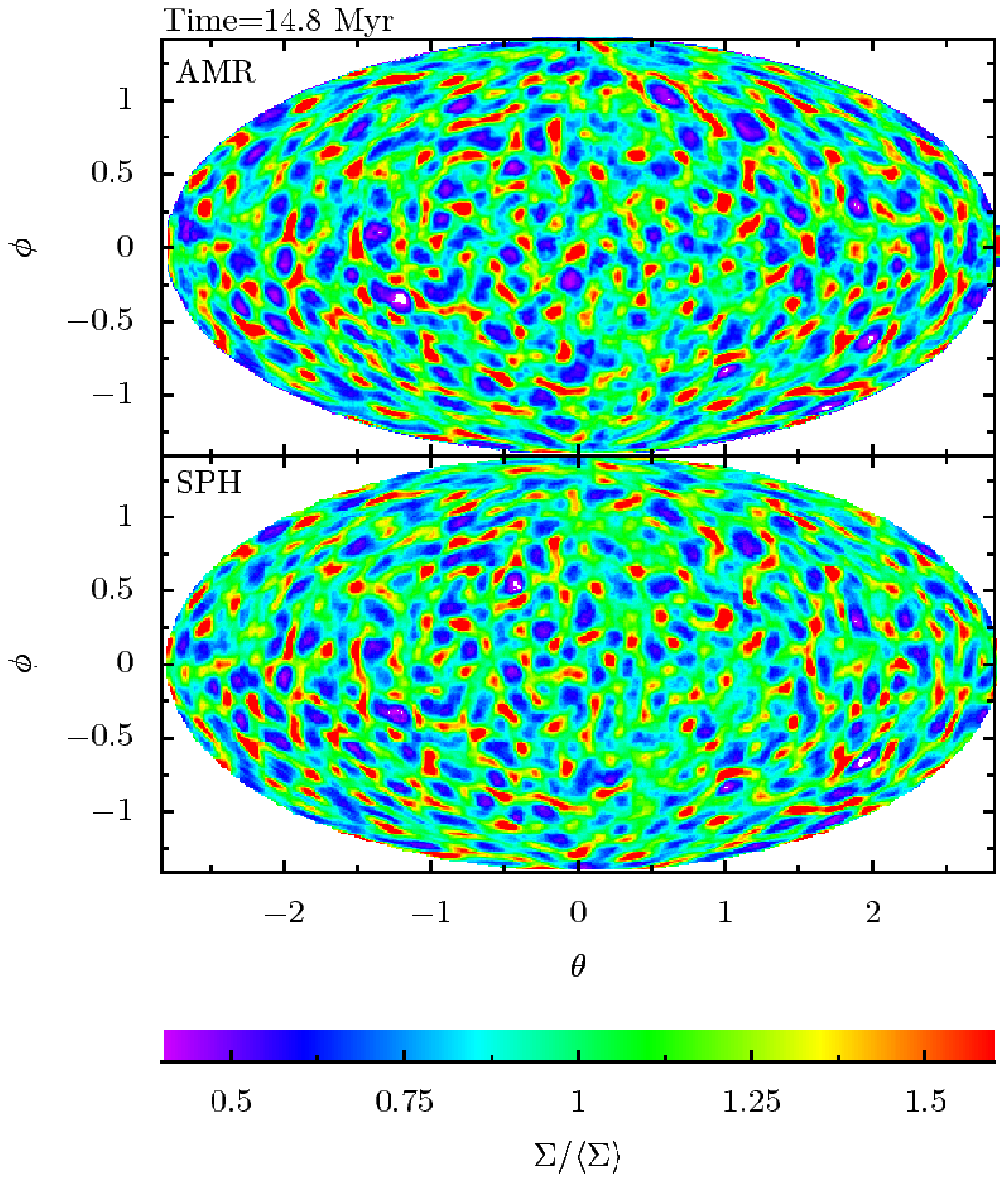}}
     \vspace{0.1in}
     \subfigure[]{\includegraphics[width=.33\textwidth]{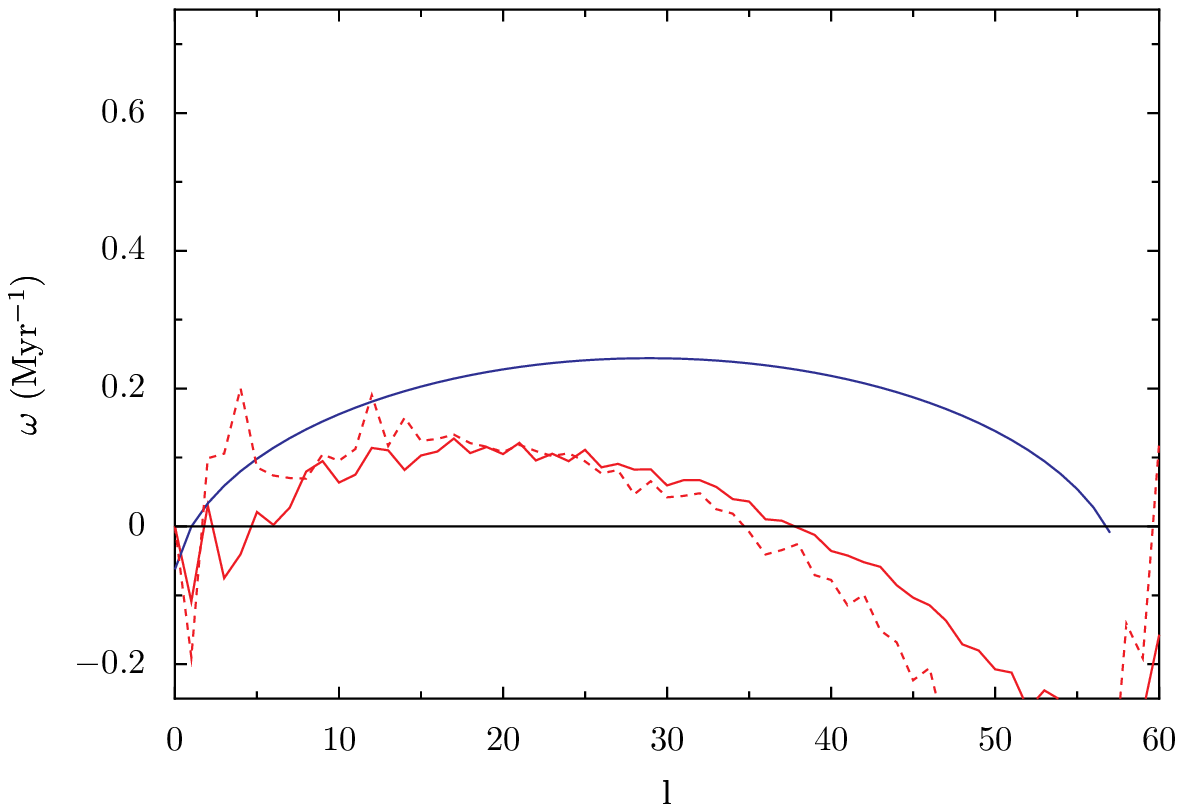}}
     \subfigure[]{\includegraphics[width=.33\textwidth]{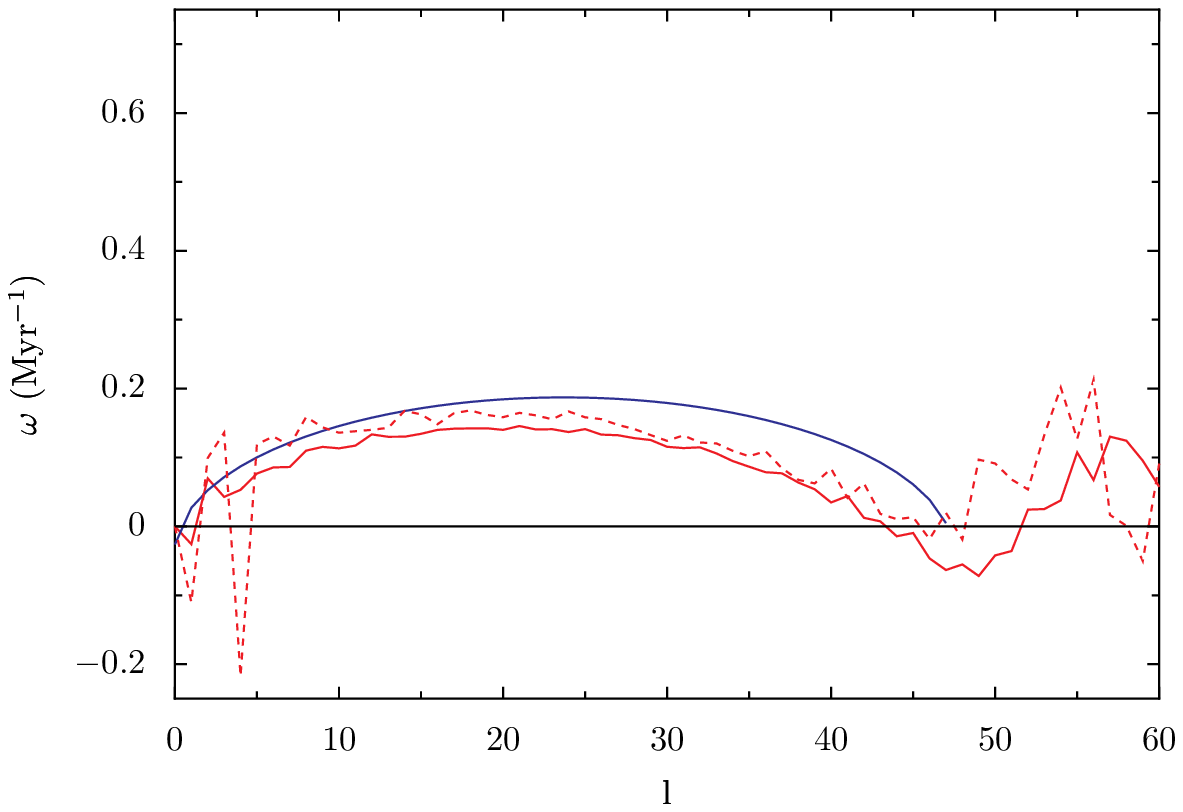}}
     \subfigure[]{\includegraphics[width=.33\textwidth]{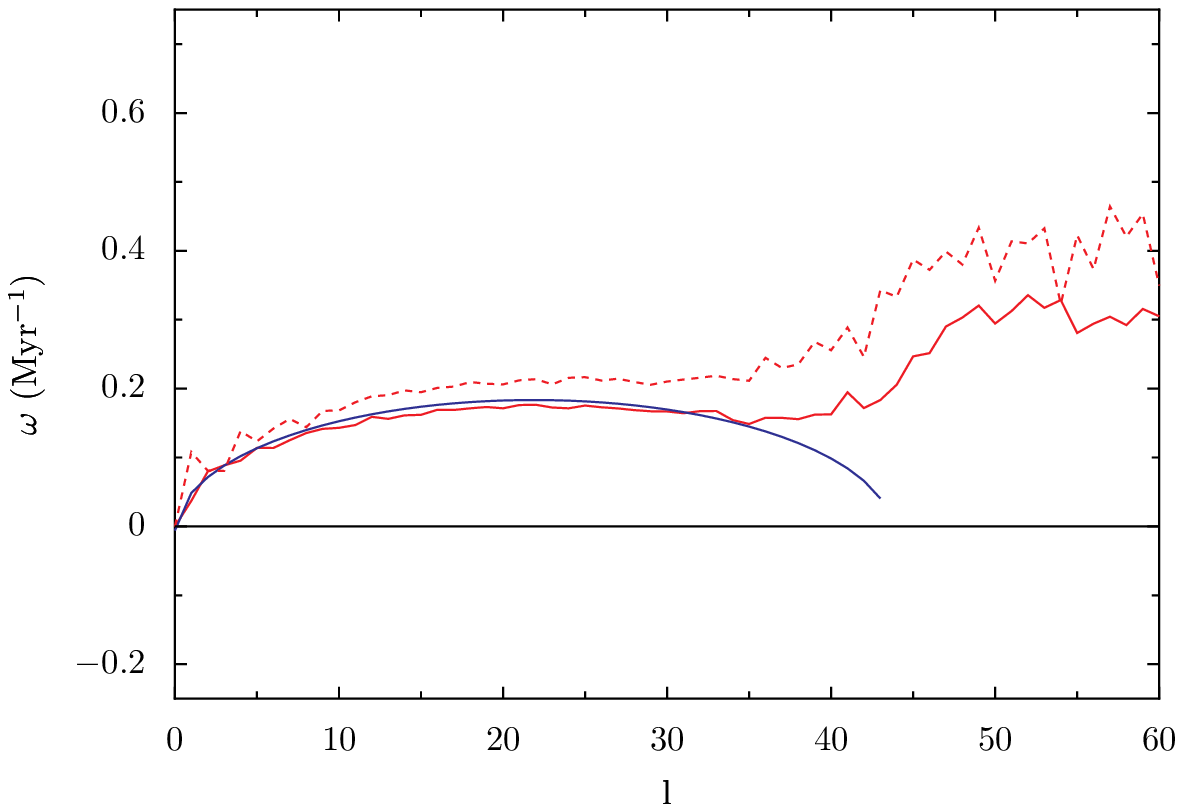}}
\caption{Hammer plots from the SPH and AMR simulations of the pressure--confined shell seeded with a Maxwellian random velocity field at three epochs (a--c) and comparisons of $\omega(l)$ derived from the SPH calculation (red solid line) and from the AMR calculation (red dotted line) with the analytical dispersion relation (solid blue line) at the same epochs (d--f). Only the real part of $\omega$ is shown.}
     \label{fig:noise_press}
\end{figure*}
\indent In these simulations, we observe again that the agreement between the two codes is very good and that there is a much closer agreement between the numerical calculations and the thin--shell approximation. The numerical dispersion relations, shown in the lower part of Figure \ref{fig:noise_press} at early times show a similar deficit in fragmentation at shorter wavelengths, but the agreement with the theoretical dispersion relation improves and, in later stages of the simulations, the simulations show fragmentation at all wavelengths predicted by the thin--shell approximation. The growth rates observed are also in good agreement with the theoretical treatment. As a result, structure in the pressure--confined shell grows much more rapidly than in the shell without a confining pressure (Figure \ref{fig:noise_nopress}). The fragmentation shows the same general form in that it appears to be filamentary in nature. It is also clear from a comparison of Figures \ref{fig:noise_nopress} and \ref{fig:noise_press} that the latter exhibits structure at smaller wavelengths, as expected from the numerical dispersion relations.\\ 
\section{Discussion}
It is clear from the results presented in Section 4 that the predictions of the thin--shell analysis as to the rates at which structures of a given wavelength will grow in a self--gravitating shell should be treated with some caution. The thin--shell approximation treats the shell as infinitesimally thin, so that the idea of the boundary conditions on the inner and outer surfaces of the shell is meaningless. Of course, no real shell is infinitesimally thin and we find that the boundary conditions applied to the shell surface affect the fragmentation very strongly. In a shell expanding in the absence of confining pressure, the radial thickening of the shell suppresses fragmentation at short wavelengths. The radial flows acting away from the middle of the shell dilute the azimuthal flows which would otherwise feed density perturbations, so that short--wavelength perturbations are unable to grow. We therefore find that the thin--shell approximation produces a rather poor description of the fragmentation occurring in such a shell. The simulations show that shells expanding without pressure confinement underproduce fragments at short wavelengths. This result is potentially interesting in an astrophysical context, since it would imply that such shells will produce a top--heavy mass function -- either a cluster mass function or a stellar IMF, depending on the parameters of the shell, in particular the total mass. Other effects, such as gas expulsion can also influence the cluster mass function and make it top--heavy, as discussed in \cite{2008ApJ...678..347P}. In either case, massive stars will be over--represented with respect to low--mass stars. This in turn makes the idea that star formation can self propagate more plausible, since triggered star formation must produce massive stars in order for this to occur.\\
\indent Conversely, we find that an expanding shell which is pressure--confined internally and externally in such a way that the thickness of the shell remains approximately constant during the shell's expansion fragments in much better agreement with the thin--shell approximation. Because the shell is prevented from thickening in the radial direction, radial flows in fact act \textit{towards} the middle of the shell and azimuthal flows dominate over radial flows in the neighbourhood of perturbations. Hence, surface density perturbations are able to grow at all spatial frequencies predicted by the thin--shell approximation at rates in good agreement with the theoretical predictions. Fragmentation occurring in this way should produce an IMF very similar in form to the Salpeter IMF \citep{2007IAUS..237..114P}. Furthermore, since the surface of the shell becomes deformed as fragments develop, the confining pressure accelerates fragmentation. Near the forming fragments, components of the pressure forces can act tangentially to the shell, driving more material into the fragments. We describe this effect as `pressure--assisted' gravitational fragmentation and will study it in detail in a subsequent paper.\\
\indent The thin--shell analysis assumes that all spatial modes evolve independently of each other, so that there is no communication or transfer of power between modes. It is not obvious that this should be true. We are able to achieve good agreement with the thin--shell analysis if the thickness of the shell is kept constant during the expansion (Figure \ref{fig:noise_press}). If mode--mode interaction implies the growth of some modes at the expense of others, there should be a corresponding change in the form of the dispersion relation. Since we do not observe this, we infer that there is no interaction of this type in our simulations.\\
\indent We have deliberately tried to eliminate both the Rayleigh--Taylor and Vishniac instabilities in our calculations by studying a shell expansion with no ram--pressure and by ensuring that there was no pressure differential across the shell. In reality, both these instabilities are likely to act on the shell and both may induce it to fragment in a manner very different to that produced by the gravitational instability acting alone. The classical Rayleigh--Taylor instability acts on all spatial lengthscales and it acts fastest on the smallest ones, so this instability is likely to generate small--scale structure in the shell much faster than the gravitational instability is able to.\\
\indent In their study of infrared shells in the Milky Way, \cite{2006ApJ...649..759C} and \cite{2007ApJ...670..428C} find a strong correlation between the radius $R$ of shells and their projected thickness $\Delta R$, such that $\Delta R\propto R$. Since shells are seen projected onto the sky, even when spherical symmetry can be assumed, the thickness measured in this way is not the same as the three--dimensional thickness of the shell. However, it is very difficult to conceive of an evolutionary scheme for the shell in which the apparent projected thickness of the shell increases but the true three--dimensional thickness does not. This result therefore implies strongly that the real thicknesses of shells increase as they expand. \cite{2006ApJ...649..759C} and \cite{2007ApJ...670..428C} also show that the shells often have intricate filamentary substructure. Based on the results presented in this paper, we suggest that these shells are likely to be undergoing gravitational fragmentation but that the thin--shell analysis will not give a reliable description of this process, given that the shells are becoming thicker as they expand. We have shown that this circumstance leads to the suppression of low--mass fragments and will result in a top--heavy IMF. The shells observed by Churchwell therefore support the notion that star formation can be self--propagating, since the production of massive stars is essential for this to occur. Observations of Sh--217 and Sh--219 \citep{2003A&A...399.1135D}, RCW 79 \citep{2006A&A...446..171Z} and Sh--104 \citep{2003A&A...408L..25D} reveal that expanding HII regions do indeed sweep up shells of dense material and trigger the formation of massive objects which in turn generate their own ultracompact HII regions and may subsequently drive further star formation. However, if such shells produce a top--heavy IMF, they cannot make a great contribution to the overall stellar population, since they do not produce an IMF consistent with that observed on large scales.\\
\indent This work is of particular relevance in the context of the self--enrichment scenario for the the formation of globular clusters. The first generation of massive stars in the core of a proto-globular cluster explode as supernovae and sweep the remaining gas into a dense shell. Since globular clusters are approximately spherical and possess smooth potentials, the shell is likely to be spherical and relatively smooth and can thus in principle be well approximated by the thin--shell approximation.\\
\indent The fate of such a shell -- whether or not it is bound and whether or not it fragments to form a second generation of stars -- is of crucial importance for several outstanding questions in globular cluster formation. It remains poorly understood why, in our galaxy, the metallicity distribution function of halo field stars extends down to $\left[{\rm Fe/H}\right]\simeq-5$, while the most metal--poor globular clusters have $\left[{\rm Fe/H}\right]\simeq-2.5$. In addition, many Galactic and some Magellanic Cloud globular clusters exhibit abundance peculiarities \citep[][and references therein]{2006PASP..118.1225S}, or multiple main--sequences \citep[e.g.][]{2007ApJ...661L..53P}. Numerous authors have suggested that these features could be explained by two or more rounds of star formation taking place within globular clusters, with the metals from each generation polluting the stars of the next. Wolf--Rayet, OB stars and AGB stars are likely to be the chief sources of pollution \citep{2006PASP..118.1225S}, so an understanding of the mass functions of later generations of stars is crucial to this model. \cite{1999A&A...352..138P} showed that it was indeed possible for globular clusters to retain supernova ejecta and the swept--up shell, and \cite{2004MNRAS.351..585P} and \cite{2005A&A...436..145R} further showed that it was possible for the shells to fragment. \cite{2004MNRAS.351..585P} used the thin--shell analysis to show that smaller numbers of first--generation supernovae and higher external pressure in the hot protogalactic background both made fragmentation of the shell more likely, since the former condition gives the shell a smaller initial velocity and the latter ensures that the shell slows down more rapidly once it leaves the protocluster cloud. As we have shown, \cite{2004MNRAS.351..585P}'s assumption of high--pressure slowing the shell down will also prevent the shell from thickening and so her results derived from the thin--shell analysis are likely to be correct. However, it may not always be the case that such shells encounter high enough pressures to keep them thin, in which case they will underproduce low mass stars.\\
\section{Conclusions}
We have conducted numerical simulations of the fragmentation of expanding gaseous shells in order to compare the numerical techniques of SPH and AMR and in order to test the theoretical predictions of the well--known thin--shell approximation.\\
\indent We find that the agreement between the SPH code and the FLASH AMR code is very good despite the differences in the numerical schemes and in the way in which constant--pressure boundary conditions are applied in the two codes. The numerically--derived dispersion relations generated by the codes agree very well.\\
\indent We find that the thin--shell approximation should be used with some caution. When dealing with the concept of an infinitesimally--thin shell, the concept of boundary conditions has no meaning, but in any finite--thickness shell, the behaviour of the shell surface must be constrained in some way. We find that the boundary conditions chosen have a strong effect on the fragmentation of the shell. In the absence of confining pressure, flows perpendicular to the shell surface cause the shell to thicken as it expands, diluting surface density perturbations. These flows preferentially wash out smaller perturbations and thus suppress fragmentation at short wavelengths. Under these conditions, the thin--shell analysis gives a poor description of the fragmentation process and the shell produces a top--heavy fragment mass function due to the deficit of small objects.\\
\indent Conversely, if a constant pressure is applied to the inner and outer surfaces of the shell such that its thickness does not change significantly as it expands, we obtain very good agreement with the thin--shell approximation in the sense that the analytical predictions of which wavelengths are unstable and which stable match the numerical computations, although we generally find that fragmentation at all wavelengths is slower than predicted by the thin--shell model.\\
\indent We note that fragmentation of the shells perturbed only by random noise (either in the density or velocity fields) results in filamentary structures. This effect has been observed in simulations of this type before \citep{2007MNRAS.375.1291D}. Although they do not mention the concept of shell fragmentation, \cite{2006ApJ...649..759C} and \cite{2007ApJ...670..428C} note that many of the infrared bubbles in their observations exhibit a similar filamentary structure.
\section{Acknowledgements}
The authors thank the anonymous refer for their very thorough reading of the paper and constructive comments.\\
The FLASH code was developed in part by the DOE-supported
Alliances Center for Astrophysical Thermonuclear Flashes at
the University of Chicago. The AMR calculations were
performed on Merlin ARCCA SRIF-3 cluster. RW acknowledges
support by the Human Resources and Mobility Programme of the
European Community under contract MEIF-CT-2006-039802. JED acknowledges support from a Marie Curie fellowship as part of the European Commision FP6 Research Training Network `Constellation' under contract MRTN--CT--2006--035890. JED, RW and JP acknowledge support from the Institutional Research Plan AV0Z10030501 of the Academy of Sciences of the Czech Republic and project LC06014--Centre for Theoretical Astrophysics of the Ministry of Educatio, Youth and Sports of the Czech Republic.

\bibliography{myrefs}

\label{lastpage}

\end{document}